\RequirePackage{amsmath}
\documentclass{iopart}

\usepackage{iopams}  
\usepackage{graphicx}
\usepackage{amssymb,amsfonts}
\usepackage{color}
\usepackage{mathrsfs}
\usepackage{cite}
\usepackage[T1]{fontenc}

\newcommand{\clr}{\color{black}}

\def\ve{\varepsilon}

\def\C{{\mathbb C}}

\def\Rve{\mathscr{R}}

\def\erfc{\mathrm{erfc}}
\def\erfcx{\mathrm{erfcx}}
\def\ctan{\mathrm{ctan}}

\begin{document}

\title{Full distribution of first exit times in the narrow escape problem}

\author{Denis S. Grebenkov}
\ead{denis.grebenkov@polytechnique.edu}
\address{Laboratoire de Physique de la Mati\`ere Condens\'ee (UMR 7643), CNRS -- Ecole Polytechnique, IP Paris, 91128 Palaiseau, France}
\author{Ralf Metzler}
\ead{rmetzler@uni-potsdam.de. Corresponding author}
\address{Institute of Physics \& Astronomy, University of Potsdam, 14476 Potsdam-Golm, Germany}
\author{Gleb Oshanin}
\ead{gleb.oshanin@gmail.com}
\address{Sorbonne Universit\'e, CNRS, Laboratoire de Physique Th\'eorique de la Mati\`ere
Condens\'ee (UMR CNRS 7600), 4 Place Jussieu, F-75005, Paris, France}
\address{Interdisciplinary Scientific Center J.-V. Poncelet (ISCP), CNRS UMI 2615, 11 Bolshoy Vlassievsky per., 119002, Moscow, Russia}

\date{\today}

\begin{abstract}
In the scenario of the narrow escape problem (NEP) a particle diffuses in a
finite container and eventually leaves it through a small "escape window"
in the otherwise impermeable boundary, once it arrives to this window and
over-passes an entropic barrier at the entrance to it. This generic problem
is mathematically identical to that of a diffusion-mediated reaction with a
partially-reactive site on the container's boundary.  Considerable knowledge
is available on the dependence of the \textit{mean\/} first-reaction time (FRT)
on the pertinent parameters. We here go a distinct step further and derive the
full FRT distribution for the NEP.  We demonstrate that \textit{typical\/}
FRTs may be orders of magnitude shorter than the \textit{mean\/} one,
thus resulting in a strong defocusing of characteristic temporal scales.
We unveil the geometry-control of the typical times, emphasising the role
of the initial distance to the target as a decisive parameter.  A crucial
finding is the further FRT defocusing due to the barrier, necessitating
repeated escape or reaction attempts interspersed with bulk excursions.
These results add new perspectives and offer a broad comprehension of various
features of the by-now classical NEP that are relevant for numerous biological
and technological systems.
\end{abstract}

%

\section{Introduction}

\begin{figure}
\begin{center}
\includegraphics[width=3cm]{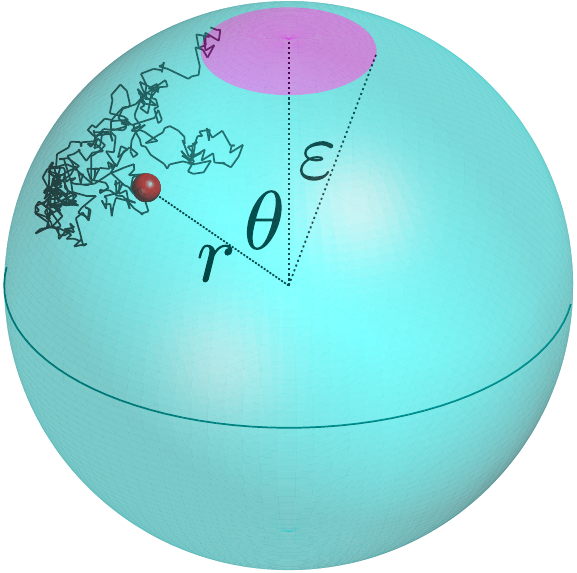}
\end{center}
\caption{Spherical domain of radius $R$ containing a target (a partially
reactive site or an EW) depicted as a spherical cap (in magenta) of a polar
angle $\ve$ located at the North pole. The red dot denotes the starting
point $(r,\theta)$ of the diffusing particle, and the zigzag line depicts
a random diffusive path to the target.}
\label{fig:1}
\end{figure}

The Narrow Escape Problem (NEP) describes the search by a diffusing
particle for a small "escape window" (EW) in the otherwise impermeable
boundary of a finite domain (figure \ref{fig:1})
\cite{Bressloff13,Holcman14}. The NEP represents a prototypical
scenario, inter alia, in biophysics, biochemistry, molecular and cell
biology, and in nanotechnology. Specifically, the particle could be an
ion, a chemically active molecule or a protein confined in a
biological cell, a cellular compartment or a microvesicle. The EW
could be a veritable "hole" in the boundary, for instance, a membrane
pore, but it could also represent a reactive site right inside (or on)
the boundary, such as a protein receptor waiting to be triggered by a
specific diffusing molecule. Similarly, the NEP may pertain to a
tracer molecule trying to leave from an interstice in the porous
structure of an underground aquifer. When one is solely interested in
how the particle reaches the target, the NEP corresponds to the
first-passage time problem \cite{Redner}. The situation becomes more
complicated when some further action is needed after the EW is
reached. For instance, the particle may have to react chemically with
an imperfect, partially-reactive site, which will require the crossing
of an activation energy barrier.  To produce a successful reaction
event the particle may repeatedly need to revisit the target after
additional rounds of bulk diffusion. Similarly, when the particle has
to leave the domain through a pore or a channel, an entropy barrier
due to the geometric confinement at the entrance to the EW needs to be
over-passed \cite{Grebenkov17}. In these scenarios, the relevant
quantities are the first-reaction or first-exit times, which are often
substantially longer than the first-passage time \emph{to\/} the
EW. For brevity, in what follows we will call the first-passage times
to the successful reaction event as the first-reaction time (FRT), for
both the eventual passage through the EW constrained by an entropic
barrier or for the reaction with a partially reactive site. This is a
random variable dependent on a variety of physical parameters and here
we unveil its non-trivial statistical properties.

The paradigmatic NEP has been extensively studied in terms of
\emph{mean\/} FRTs \cite{Holcman14,Grigoriev02,Metzler,Benichou14}.
While early analyses concentrated on idealised, spherically-symmetric
domains enclosed by a perfect hard-wall with a perfect (that is,
barrierless) EW \cite{Rayleigh,Holcman04,Singer06a,Schuss07,Cheviakov10,
Benichou08,Caginalp12}, more recent work examined the NEP for domains with
more complicated geometries and non-smooth boundaries, when, e.g., a perfect
barrierless target is hidden in a cusp or screened by surface irregularities
\cite{Singer06c,Pillay10,Cheviakov12,Grebenkov16,Grebenkov19}.  A striking
example of the NEP with very severe escape restrictions is the dire strait
problem \cite{dire}.  Effects of short-ranged (contact) and long-ranged
interactions with the confining boundary, which are quite ubiquitous and give
rise to "intermittent motion" characterised by alternating phases of bulk and
surface diffusion, were shown to allow for an optimisation of the search times
\cite{Oshanin10,Benichou10,Benichou11,Rupprecht12,Grebenkov17,Ber12a,Ber12b}.
It was also studied how molecular crowding in cellular environments
impacts the search dynamics \cite{Benichou08,Agranov18,Lanoiselee18}.
When the target is imperfect the contributions due to diffusive search
and penetration through a barrier were shown to enter additively to the
\emph{mean\/} FRT \cite{Rojo12,Grebenkov17,Grebenkov17a,Grebenkov19d}, akin to the
celebrated Collins-Kimball relation in chemical kinetics
\cite{Collins49}.  Since these two rate-controlling factors
disentangle the concept of search- versus barrier-controlled NEP was
proposed \cite{Grebenkov17,Grebenkov17a}.

Despite of this considerable body of works published on the NEP, information
beyond the \emph{mean\/} FRT is scarce, only two recent analytical
and numerical works consider the full distributions of first-passage
times to \emph{perfect\/} targets and only in 
two-dimensional settings \cite{Rupprecht15,Hafner18}.  In fact, the NEP
typically involves a broad spectrum of time scales giving rise to a rich
and interesting structure of the probability density function (PDF)
of FRTs with different time regimes.  Knowledge of this full PDF is therefore
a pressing case. As we show here the FRT may be strongly defocused over
several orders of magnitude, effecting noticeable trajectory-to-trajectory
fluctuations \cite{Mejia12}. One therefore cannot expect that, in general,
solely the first moment of the PDF can be sufficient to characterise its form
exhaustively well. Moreover, it is quite common that the behaviour of the
positive moments of a distribution is dominated by its long-time tail and,
hence, stems from anomalously long and rarely observed trajectories
\cite{Mejia11,Mejia12,Godec16}. Fluctuations between realisations are indeed
common in diffusive processes  and can be characterised by the
amplitude fluctuations of time-averaged moments \cite{he,johannes} and
single-trajectory power spectra \cite{glebps,glebps1,vittoria}.

We here present the PDF of the FRTs for the NEP in the generic case
of a spherical domain with an \textit{imperfect\/} target and an arbitrary
{\it fixed} starting point of the diffusing particle.  Indeed, in cellular
environments, this setting  represents one
of the most interesting applications of the NEP, as particles such as
small signalling molecules or proteins are usually released at some fixed
point inside the cell \cite{xie} and need to diffusively locate specific
targets such as channel or receptor proteins embedded in the cell membrane
\cite{bio,bio1}.  Our analytical approach is based on a self-consistent
closure scheme \cite{Shoup81,Grebenkov17,Grebenkov18} that yields analytical
results in excellent agreement with numerical solutions.  We demonstrate that
in these settings the PDF exhibits typically \emph{four distinct temporal
regimes} delimited by three relevant time scales, for which we also present
explicit expressions.  These characteristic times may differ from each
other by several orders of magnitude.  We fully characterise the PDF of
FRTs by identifying the cumulative reaction depths (the fractions of FRT
events corresponding to each temporal regime of the PDF) and analyse their
dependence on the system parameters.  Overall, our analysis provides a first
complete and comprehensive framework for the understanding of multiple facets
of the biologically and technologically relevant NEP, paving a way for a
better theoretical understanding of the NEP and a meaningful interpretation
of experimental data. Last but not least, it permits for a straightforward
derivation of the first-reaction time PDF for a more general problem with
several diffusive particles starting their random motion from given locations.
In appendices, we also discuss the forms of the PDF and the relevant time scales
for other possible initial conditions.

\section{The Narrow Escape Problem}

As depicted in figure \ref{fig:1} we consider a diffusing point-like particle
with diffusion coefficient $D$ inside a spherical domain of radius $R$.
The boundary enclosing the domain is perfectly reflecting everywhere, except
for the \textit{imperfect\/} target -- a spherical cap of angle $\ve$ at the
North pole.  By symmetry, the behaviour is independent of the azimuthal angle,
and we need to solve the diffusion equation for the survival probability
$S(r,\theta;t)$ that a particle released at position ${\bf r}=(r,\theta)$
at time $0$ has not reacted with the partially-reactive site or has not
escaped through the EW up to time $t$.  The diffusion equation is completed
by the initial condition $S(r,\theta; t=0) = 1$ and by the {\it mixed\/} boundary conditions of
zero-current at the hard wall combined with the standard Collins-Kimball
partially-reactive (or partially-reflecting) boundary condition imposed
at the EW or the chemically-active site \cite{Collins49} (see also
\cite{Grebenkov19b,gleb} for an overview)
\begin{equation}  \label{eq:mixed}
- \left. D\frac{\partial S(r,\theta;t)}{\partial r}\right|_{r=R}=
\left\{ \begin{array}{ll} \kappa \, S(R, \theta ;t) & (0 \leq \theta < \ve), \\
0 & (\ve \leq \theta \leq \pi). \\ \end{array} \right. 
\end{equation} 
The latter Robin boundary condition signifies that the reaction is a
two-stage process consisting of a) the diffusive transport of the
particle to the vicinity of the target and b) the eventual imperfect
reaction which happens with a finite probability.  The proportionality
factor $\kappa$ here is the intrinsic reactivity, which is defined as
$\kappa = \omega l$, where $l$ is the effective "thickness" of the
reaction zone in the vicinity of the target -- the minimal reaction
distance, while $\omega$ is the rate (frequency) describing the number
of elementary reaction acts per unit of time within the volume of the
reaction zone. Clearly, $\omega$ (and hence, $\kappa$) depends on the
amplitude of the barrier against the reaction event: $\omega = \infty$
($\kappa= \infty$) corresponds to the case of a perfect barrierless
reaction or an immediate escape upon the first encounter with the EW.
In this case the FRT reduces to the first-passage time to the target.
Conversely, $\omega = 0$ ($\kappa= 0$) implies that the
reaction/escape event is completely suppressed. In our calculations,
we suppose that $\kappa$ is a given parameter (for
heterogeneous, space-dependent reactivity, see \cite{Grebenkov19c} and
references therein).  In Section \ref{sec:discussion}, we will
elaborate on possible distinctions between its values for reaction and
escape processes.  Once $S(r,\theta;t)$ is obtained, the PDF
$H(r,\theta;t)$ of the FRT is given by the negative derivative of the
survival probability with respect to $t$.  Further details on the NEP,
our analytical derivation, and the comparison with numerical results
are presented in appendices.

\begin{figure*}
\centering
\includegraphics[width=\textwidth]{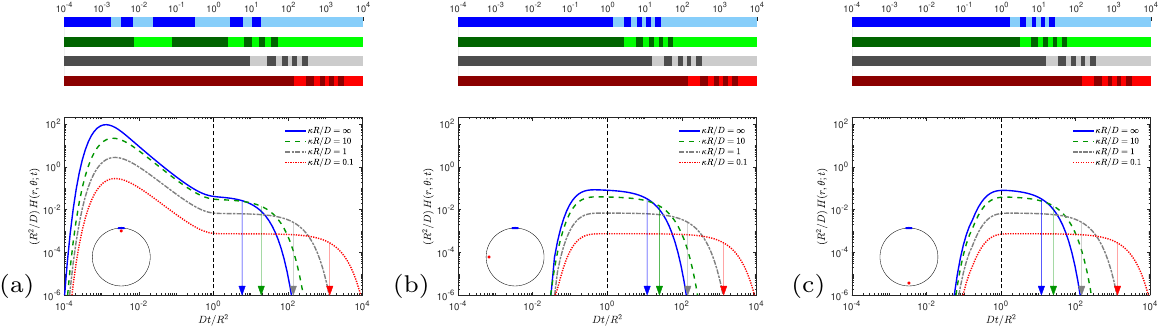}
\caption{PDF $H(r,\theta;t)$ of FRTs, from Eq. \eqref{chief}, rescaled by $R^2/D$,
as function of $D t/R^2$ for $r/R=0.9$, $\ve=0.1$, four values of the
dimensionless reactivity $\bar{\kappa}=\kappa R/D$ indicated in the
panels, and {\bf (a)} $\theta=0$, {\bf (b)} $\theta=\pi/2$, {\bf (c)}
$\theta=\pi$.  Insets in all panels indicate schematically the
location of the starting point shown by red circle, with respect to
the target (blue) placed at the North pole.  Note that
$\bar{\kappa}$ includes $R$ and $D$ such that at fixed $\kappa$,
smaller values of $\bar{\kappa}$ can be achieved by lowering $R$ or
increasing $D$.  The  approximation in \eqref{chief} is truncated to
$n=50$ terms.  Vertical arrows indicate the mean FRT $t_{\rm mean}$.
The dashed vertical line shows the crossover time $t_c =R^2/D$, {at
which the particle first engages with the boundaries}.  The coloured
bar-codes above each main panel indicate the cumulative reaction
depths corresponding to the four values of $\bar{\kappa}$, in
decreasing order from top to bottom.  Each bar-code is split into ten
regions of alternating brightness, representing the ten
$10\%$-quantiles of the PDF.  The numbers on top of the bar-codes
indicate the values of $D t/R^2$.  For example, in panel {\bf (a)} the
first dark blue region for the case $\bar{\kappa}=\infty$ (perfect
reaction or no entropic barrier at the EW) indicates that $10\%$ of
reaction events occur up to $Dt/R^2\approx3\times10^{-3}$, which is
close to the dimensionless most probable FRT $D t_{\rm mp}/R^2$.  The
mean FRT $t_{\rm mean}$ in this case is almost four orders of
magnitude longer than the most probable FRT $t_{\rm mp}$, and over
$70\%$ of trajectories arrive to the target up to this time. Analogous
estimates of the cumulative reaction depths for other initial
conditions are presented in the text below and also in appendices.}
\label{fig:fig22}
\end{figure*}

\section{Results}

Our principal analytical result is the explicit,  approximate expression for the
PDF $H(r,\theta;t)$ of the FRT,
\begin{equation}
\label{chief}
H(r,\theta;t)=\frac{D}{R^2}\sum\limits_{m=0}^\infty e^{-Dt\alpha_m^2/R^2}v_m(
r,\theta),
\end{equation}
with
\begin{eqnarray}
&&v_m(r,\theta)=\frac{2}{F'(\alpha_m)}\sum\limits_{n=0}^\infty\phi_n(\ve)P_n(
\cos\theta)\frac{j_n(\alpha_m r/R)}{-j'_n(\alpha_m)},\\
&&\phi_n(\ve)=\frac{P_{n-1}(\cos\ve)-P_{n+1}(\cos\ve)}{1-\cos\ve}\quad(n=1,2,
\ldots),
\end{eqnarray}
where $\phi_0(\ve)=1$, $P_n\left(\cos \ve\right)$ are the Legendre polynomials,
$j_n(z)$ are the spherical Bessel functions of the first kind, the prime
denotes the derivative with respect to the argument, while $\alpha_m$ are
the positive solutions (organised in ascending order) of the transcendental
equation
\begin{equation}
\label{F}
\sum_{n=0}^{\infty} \frac{\phi_n^2\left(\ve\right)}{\left(2 n + 1\right)} 
\frac{j_n(\alpha)}{\alpha j'_n(\alpha)}\equiv F(\alpha)=-\frac{2 D}{\kappa
R\left(1-\cos\ve\right)}.
\end{equation}
Equations \eqref{chief} to \eqref{F} completely determine the explicit
form of $H(r,\theta;t)$. {\clr As we already mentioned above, this result
is obtained by resorting to a self-consistent closure scheme, developed
earlier for the calculation of the mean FRT in certain reaction-diffusion
problems \cite{Shoup81,Grebenkov17,Grebenkov18}. This approximation consists
in replacing the actual mixed boundary condition \eqref{eq:mixed} by an
inhomogeneous Neumann condition and in the derivation of an appropriate closure
relation, which ensures that the mixed boundary condition \eqref{eq:mixed}
holds {\it on average}. The adaptation of such a scheme to the calculation
of the full PDF in NEP is discussed in detail in \ref{apA} and \ref{apB}.}
We note that despite an approximate character of our approach,
the obtained result turns out to be very accurate for arbitrary starting
position $(r,\theta)$ (although not too close to the target), for a very
wide range of $\varepsilon$, and for arbitrary value of $\kappa$.  {\clr The
accuracy was checked by comparing our approximation to a numerical solution
of the original problem by a finite-elements method (see \ref{sec:FEM}).}
We note, as well, that two important characteristic length-scales
can be read off directly from expression \eqref{chief} and are associated with
the smallest and the second smallest solutions of \eqref{F}, (that being,
$\alpha_0$ and $\alpha_1$), which define the largest and the second largest
decay times of the PDF, respectively. The analysis of all the solutions of
\eqref{F}, which define the full form of the PDF valid at arbitrary times,
their relation to the characteristic time scales, the derivations of the
asymptotic forms of $H(r,\theta;t)$, and its easy-to-implement approximate
mathematical form are provided in the appendices. The latter also present
a variety of other useful results based on $H(r,\theta;t)$, e.g., the
surface-averaged and volume-averaged PDFs (see Eqs. \eqref{eq:Ht_volume}
and \eqref{eq:Ht_surface}).

In figure \ref{fig:fig22} we depict the PDF $H(r,\theta;t)$ defined by
\eqref{chief} to \eqref{F} for three different fixed initial positions
(polar angles  $\theta = 0$ (a), $\theta = \pi/2$ (b), and $\theta = \pi$ (c), with fixed
$r/R = 0.9$), four values of the intrinsic reactivity
$\kappa$, and for a particular choice $\ve = 0.1$.  We observe that the PDF
has a clearly defined structure, which depends on whether the starting
point is close to the target or not.  In the former case (as in panel (a)),
the PDF consists of a hump-like region around the most probable FRT $t_{\rm
mp}$, an extended plateau-like region after the crossover time $t_{\rm c}$
(remarkably, within this region all values of the FRT are nearly equally
probable), and, ultimately, exponential decay starting right after the
mean FRT $t_{\rm mean}$. Note that up to $t_{\rm mean}$
the particle experiences multiple collisions with the boundary of the
container after unsuccessful reactions, thereby loosing the memory about
its precise starting location. Due to this circumstance, the ultimate,
long-time exponential decay (which is the most trivial part of the PDF),
holds for all bounded domains, regardless of the precise initial condition
(see also appendices). As evidenced in figure \ref{fig:fig22} the difference between
$t_{\rm mp}$, $t_{\rm c}$ and $t_{\rm mean}$ may span orders of magnitude
revealing a pronounced defocusing of the FRTs (see below).  In contrast,
when the starting point is far from the target (panels (b) and (c)), the
hump-like region is much less apparent, and the extended plateau and the
exponential decay are the major features of the PDF. Although the maximum
of the PDF formally exists, the corresponding most probable time is not
informative because the escape (or reaction) may happen at any moment over
the plateau region with almost equal probabilities. We note that the overall shape of the curves
presented in figure \ref{fig:fig22} is generic and gets only slightly modified
upon parameter changes -- see, e.g., figure \ref{fig:fig2} in appendices in which a
similar plot for $r/R = 0.5$ is presented.  Note, as well, that the behaviour
of the PDF in the case when the starting point is uniformly distributed over a
spherical surface some distance $r$ away from the origin, appears to be very
similar to the one depicted in figure \ref{fig:fig22} in panels (b) and (c)
(see figure \ref{fig:Ht_r05}, panel (b)).  Conversely, in the case when the
location of the starting point is random and uniformly distributed over the
volume of the container, the PDF consists essentially of a plateau-like region
and the ultimate exponential tail (see figure \ref{fig:Ht_r05} (a)).

The existence of the hump-like region is not an unexpected feature in
the case of a fixed initial condition.  For a particle initially
placed some distance away from the target, it is impossible to reach
the latter instantaneously such that the probability of having a very
short FRT is close to zero. This defines the most probable reaction
time $t_{\rm mp}$, which can be estimated as function of the initial
distance $\sigma$ to the target (in panel (a) of Fig. \ref{fig:fig22},
$\sigma=R-r$ as $\theta=0$) in the form (see appendices),
\begin{equation}
t_{\rm mp}\lesssim(R-r)^2/(2D).
\end{equation}
The most probable FTR corresponds to "direct" trajectories
\cite{aljaz} on which the particle moves fairly straight to the
target, with immediate successful reaction. Consequently the value of
$t_{\rm mp}$ is "geometry-controlled" \cite{aljaz,Grebenkov18b} by the
initial distance $R-r$.  At such short time scales, the particle has
not yet explored the boundaries of the domain, and thus the initial
increase up to $t_{\rm mp}$ and the subsequent power-law descent of
the PDF are well-described by the L\'evy-Smirnov-type law $A
\exp(-\alpha/t)/t^{3/2}$ \cite{Redner} (see \ref{sec:Ashort} for more
details). In our case of an imperfect target, the amplitude $A$ also
acquires the dependence on the reactivity $\kappa$ and the target
radius $\rho$. In particular, the descent from the peak value of the
PDF obeys
\begin{equation}
\label{eq:Levy-Smirnov}
H(r,\theta;t)\simeq\frac{\rho}{\sqrt{4\pi Dt^3}}\frac{\frac{D}{\kappa\rho}+
\frac{\sigma}{\sigma+\rho}}{(1+\frac{D}{\kappa\rho})^2} \,.
\end{equation}
The independence on the domain size stems from the fact that the particle
at such short times has no knowledge about the boundaries.

The hump-like region is delimited by the crossover time
\begin{equation}
t_{\rm c}\approx R^2/D,
\end{equation}
an important characteristic time-scale, at which the particle first
engages with the domain boundaries and realises that it lives in a bounded domain. 
The plateau-like regime, which sets
in at $t_{\rm c}$, is a remarkable feature for the NEP that has not been
reported in literature on this important and generic problem.  The existence
of such a plateau for diffusion-reaction systems with an activation energy
barrier at the target has only been recently discovered in other geometrical
settings, such as, e.g., the diffusive search for a partially-reactive
site placed on a hard-wall cylinder confined in a cylindrical container
\cite{Grebenkov18} and a spherical target at the centre of a larger spherical
domain \cite{Grebenkov18b}. The fact that it also appears in typical NEP
settings, as shown here, evidences that it is a fundamental feature of
systems in which the reaction event requires not only a diffusive search
but also a barrier crossing.

The plateau typically persists for several decades in time, up to the mean FRT
$t_{\rm mean}$ (indicated by vertical arrows in figure \ref{fig:fig22}), which
is very close (but not exactly equal) to the decay-time $t_{\rm
d}= 1/(D\lambda_0)$, associated with the smallest eigenvalue $\lambda_0$
of the Laplace operator (see appendices for more details). As a consequence,
the FRT PDF features the exponential shoulder
\begin{equation}
H(r,\theta;t) \sim \exp\left(- t/t_{\rm d} \right).
\end{equation}
Note that the right-tail of the PDF is thus fully characterised by the unique
time scale $t_{\rm d}$. For a perfectly-reactive target, the asymptotic
behaviour of $t_{\rm d}$ in the narrow-escape limit (i.e., for $\ve\to0$)
was first obtained by Lord Rayleigh \cite{Rayleigh} more than a
century ago within the context of the theory of sound, and then refined
and adapted for the NEP by Singer {\it et al.} \cite{Singer06a}, yielding
(see also discussions in appendices and in \cite{Ward93,Isaacson13,Grebenkov13})
\begin{equation}
t_{\rm d}\simeq\frac{3R^2}{\pi D}\ve^{-1}\qquad(\ve\to0).
\end{equation}
For an imperfect target, however, this asymptotic result
is no longer valid and one finds instead that (see appendices)
\begin{equation}
\label{d}
t_{\rm d}\simeq\frac{2R}{3\kappa(1-\cos\ve)}\simeq\frac{4R}{3\kappa}\ve^{-2}
\qquad(\ve\to0)
\end{equation}
which signifies that the decay time diverges much {\it faster\/} when
$\ve\to0$, as compared to the case of a perfect barrierless
target. Lastly, we note that the fact that $t_{\rm mean}$ and $t_{\rm
d}$ appear very close to each other is an apparent consequence of the
existence of the plateau region with an exponential cut-off,
suggesting that the major contribution to $t_{\rm mean}$ (and most
likely to all positive moments of the PDF) is dominated by the
integration over this temporal regime.

\section{Discussion}
\label{sec:discussion}

The results presented above reveal many interesting and novel features of
the NEP that we now put into  more general context.

i) In case of chemical reactions with a partially-reactive site, the
intrinsic reactivity $\kappa$ is independent of $\ve$.  As a consequence,
$t_{\rm d} $ in Eq. (\ref{d}) (as well as $t_{\rm mean}$, see appendices)
diverges in the NEP limit as $t_{\rm d} \sim 1/\ve^{2}$.  The situation
is much more delicate in the case of an entropy barrier $\Delta S$. The
impeding effect of the latter has been studied in various guises,
including transport in narrow channels with corrugated walls (see, e.g.,
\cite{channel0,channel1,channel2,channel3,channel4}) and also for the NEP (see,
e.g., \cite{escape0,escape1}), although for the latter problem no explicit
results have been presented.  In particular, for diffusion in channels
represented as a periodic sequence of broad chambers and bottlenecks, the
influential works \cite{channel0,channel2} expressed this barrier through a
profile $h(x)$ of the channel cross-section, while in \cite{channel4}, for
ratchet-like channels, this barrier was estimated as $\Delta S \sim \beta^{-1}
\ln\left(h_{\rm max}/h_{\rm min}\right)$, where $\beta$ is the reciprocal
temperature, $h_{\rm max}$ and $h_{\rm min}$ are the maximal and minimal
channel apertures (a chamber versus a bottleneck). Therefore, one may expect
that for the NEP the entropic barrier scales as $\Delta S \sim \beta^{-1}
\alpha \ln\left(1/\ve\right)$, where $\alpha > 0$ is a numerical constant
characterising the precise form of the profile
$h(x)$.  Assuming that the particle penetrates through the barrier solely
due to a thermal activation, i.e., that $\kappa \sim \exp(-\beta \Delta S)$,
one expects then that $\kappa \sim \ve^{\alpha}$ and thus vanishes in the NEP
limit $\ve \to 0$. This means, in turn, that here $t_{\rm d} \sim 1/\ve^{2 +
\alpha}$, i.e., $t_{\rm d}$ diverges more strongly in the NEP limit than its
counterpart in the case of a partially-reactive site.  As a consequence,
the overall effect of such an entropy barrier on the size of the plateau
region and on the cumulative reaction depth on this temporal stage, is much
more significant for the NEP than for diffusion-reaction systems. Importantly,
both for the NEP and for reactions with a partially-reactive site, as $\ve \to
0$, the mean FRT is always controlled by the penetration through the barrier
rather than by the stage of random search for the target whose contribution
to $t_{\rm d}$ diverges in the leading order only as $1/\ve$.

We also remark that, of course, the NEP limit has to be understood with an
appropriate caution: taking $\ve \to 0$, one descends to molecular scales
at which both the molecular structure of the container's boundaries, surface
irregularities, presence of contaminants and other chemically active species
come into play. At such scales, the solvent can also be structured close to
the wall, affecting characteristic diffusion times and hence, the value of the
diffusion coefficient $D$.  Moreover, the size of a particle does matter here
and one can no longer consider it as point-like. The entropic barrier is then
a function of the ratio between the particle radius $a$ and the radius $\rho$
of the EW. Understanding theoretically the functional form of $\kappa$ is so
far beyond reach and one may address the problem only by numerical simulations
or experiments. We are, however, unaware of any systematic approach
to these issues, except for a recent paper \cite{chinese}, which studied
the particle-size effect on the entropic barriers via numerical simulations.
It was observed that the dependence of the inverse transition rate (and hence,
of $\kappa$) of a Brownian particle in a model system consisting of a necklace
of hard-wall spheres connected to each other by EWs, can be well-fitted
by an empirical law of the form $\kappa \sim \left(1- a/\rho\right)^{3/2}$
signifying that $\kappa$ vanishes very rapidly when $a \to \rho$. When the
particle size $a$ is comparable to that of the EW, the entropic barrier
becomes very high resulting in a significant decrease of $\kappa$.  
{\clr Some other effects of the particle size and shape onto diffusive search,
unrelated to our work, were discussed in \cite{Holcman12,Malgaretti12}.}

ii) The concept of the search-controlled versus barrier-controlled NEP
proposed in \cite{Grebenkov17,Grebenkov17a} holds indeed for the mean FRT,
see Eq. \eqref{m}. In contrast, we observe that two other important characteristic
time-scales -- $t_{\rm mp}$ and $t_{\rm c}$ -- are unaffected by the
reactivity and solely depend on the diffusion coefficient and the geometry.
This means that the two controlling factors represented by $\kappa$ and
$D$ do not disentangle. Moreover, both affect the shape of the PDF and the
corresponding cumulative reaction depths.

iii) For the NEP with fixed starting point of the particle the characteristic
time-scales $t_{\rm mp}$, $t_{\rm c}$ and $t_{\rm mean}$ differ by orders
of magnitude meaning that the FRT are typically very defocused.  This is a
conceptually important point for the understanding of the NEP, but alone
it does not provide a complete picture. Indeed, one should also analyse
the cumulative reaction depths corresponding to each temporal regime.
For example, for the settings used in figure \ref{fig:fig22} (panel (a)),
the shortest relevant time-scale is $t_{\rm mp}$.  For perfect, barrierless
reactions the amount of reaction events appearing up to this quite short
time $(D t/R^2 \approx 10^{-3})$ is surprisingly non negligible, being of the
order of $10$ per cent.  Upon lowering the dimensionless reactivity $\bar{\kappa}
= \kappa R/D$ (i.e., upon increase of the activation barrier at the target or
of the diffusion coefficient $D$, or upon lowering $R$), this amount drops
significantly and the contribution of "direct" trajectories becomes less
important.  This is quite expected because the latter now very rarely lead to
the reaction event.  In turn, the contribution of the whole hump-like region
(for $t$ stretching up to $t_{\rm c}$) for barrierless reactions is more
than $50$ per cent, meaning that more than a half of all reaction events take
place up to the time-scale when the particle first realises that it moves in
a confined domain.  One infers next from figure \ref{fig:fig22} (panel (a))
that upon lowering $\bar{\kappa}$, this amount also drops rapidly.  At the same
time, the plateau-like region becomes increasingly more pronounced for smaller
$\bar{\kappa}$ and progressively more reaction events take place during this stage:
we find $25$, $30$, $47$ and $55$ per cent of all reaction events for $\bar{\kappa}
= \infty$, $10$, $1$ and $0.1$, respectively.  Interestingly, the fraction of
"unsuccessful trajectories", which survive up to the largest characteristic
time $t_{\rm mean}$, appears to be very weakly dependent on the value of
$\bar{\kappa}$, and is rather universally about $30$ per cent. Note that this
number is close to the fraction of outcomes of an exponentially distributed
random variable above its mean, $\exp(-1) \simeq 0.37$. Therefore, even for
this particular case when the starting point is very close to the target, one
observes a rather diverse behaviour -- there is no apparent unique time-scale
and no overall dominant temporal regime: the region with the rise to the
most probable value, the one with the subsequent descent to the plateau,
the plateau-like region itself, and the eventual exponential decay may all
become relevant for different values of the parameters and hence, their
respective contributions depend on the case at hand.  This illustrates our
earlier statement that the knowledge of the full PDF of FRT is indispensable
for the proper understanding of distinct facets of the NEP.

For a random starting point, when it is either uniformly distributed over
the volume, or is located at a random point on a distance $r$
away from the origin, the situation is somewhat different. Here, the hump-like
region is smeared away and the PDF consists of just two regions:
a plateau-like region and the ultimate exponential decay. As a consequence,
in these cases $t_{\rm mean}$ is a unique characteristic time-scale, as it
was claimed previously (see, e.g. \cite{Benichou14}).  In appendices, we present
explicit results for the general case of diffusion-mediated reactions with an
activation energy barrier and imperfect EW with an entropy barrier. Concerning
the cumulative reaction depths in case of a random starting point, we observe
an analogous surprising behaviour to what was noted above: the cumulative
reaction depth over the plateau region (up to $t_{\rm mean}$) appears to
be rather universally (i.e., independent of $\kappa$ and other parameters)
equal to $\approx70$ per cent, while the remaining $\approx30$ per cent of
trajectories react within the final exponential stage.

iv) Apart from the analysis of the cumulative depths, it is helpful to work
out an explicit approximate formula for the PDF, which may be used to fit
experimental or numerical data. In appendices, we outline a derivation of such
a formula, which agrees very well for the whole range of variation of the FRT.


v) One often deals with situations when several diffusing
particles, starting at some fixed points, are present in a biophysical
system and the desired random event is triggered by the one, which
arrives {\it first} to a specific site on the surface of the
container. Among many examples (see e.g. \cite{holca}), one may
mention, for instance, calcium ions which activate calcium release in
the endoplasmic reticulum at neuronal synapses. Our results for the
fixed starting-point PDF $H(r,\theta;t)$ and the corresponding
survival probability provide the desired solution to this problem
too. As a matter of fact, if the particles are present at a nanomolar
concentration, such that their dynamics are independent, the
probability that neither of the particles arrived to the target site
up to time $t$ is just the product of the single-particle survival
probabilities. The PDF of the time of the first reaction event then
follows by a mere differentiation of this product.

\section{Conclusion}
\label{disc}

With modern optical microscopy, production events of single proteins
\cite{xie}, their passage through the cell \cite{gratton} as well as
single-molecule binding events \cite{mazza,xie1} can be monitored in
living biological cells \cite{elf}.  Typical processes in cells such as
gene expression involve the production of proteins at a specific location,
their diffusive search for their designated binding site, and the ultimate
reaction with this partially-reactive site \cite{mirny,otto}.  Given the
often minute, nanomolar concentrations of specific signalling molecules in
cells, the first-reaction time PDF of a diffusing molecule with its target in
the cell volume was shown to be strongly defocused and geometry-controlled
\cite{aljaz,Grebenkov18,Grebenkov18b}.  These analyses demonstrate that
the concept of mean FRTs (or reaction rates) is no longer adequate in such
settings involving very low concentrations.  Here one invariably needs
knowledge of the full PDF to describe such systems faithfully.

Complementary to this situation when a particle is released inside a finite
volume and needs to react with a target somewhere else in this volume, we here
analysed the important case when a diffusing particle is released inside a
bounded volume and needs to react with a target on the boundary. Our results
for the FRT PDF including the passage of an energetic or entropic barrier
leading to final reaction with the target or the crossing of the escape window,
demonstrate a pronounced defocusing of time scales.  While the most likely
FRT corresponds to a direct particle trajectory to the target and immediate
reaction, indirect paths decorrelating the initial position by interaction
with the boundary are further emphasised by unsuccessful reaction events,
leading to further rounds of exploration of the bulk.  This effect is shown
to imply a pronounced defocusing, and thus a large reaction depth is reached
long before the mean FRT. In particular, we demonstrate that the shapes of the
PDF are distinctly different from the case when the particle needs to react
with a target inside the volume of the bounded domain.

Our findings have immediate relevance to biological cells, in which specific
molecules released in the cell need to locate and bind to receptors on the
inside of the cell wall, or they need to pass the cell wall or compartment
membrane through protein channels or membrane pores.  Similar effects occur in
inanimate systems, when chemicals in porous aquifers need to leave interstices
through narrow holes in order to penetrate further.  From a technological
perspective our results will find applications in setups involving lipid
bilayer micro-reaction containers linked by tubes \cite{orwar}.

\section*{Acknowledgments}
The authors acknowledge stimulating discussions with A. M. Berezhkovskii.
DG acknowledges the support under Grant No. ANR-13-JSV5-0006-01 of the
French National Research Agency. RM acknowledges DFG grant ME 1535/7-1 from the
Deutsche Forschungsgemeinschaft (DFG) as well as the Foundation for Polish Science
for a Humboldt Polish Honorary Research Scholarship.







\clearpage
\appendix
\section{Solution in Laplace domain}
\label{apA}

Our starting point is the diffusion equation for the survival
probability $S(r,\theta;t)$ of a particle started at position
$(r,\theta)$ inside a sphere of radius $R$, up to time $t$:
\begin{equation}
\partial_t S(r,\theta;t) = D \Delta S_t(r,\theta),
\end{equation}
subject to the initial condition $S(r,\theta; t=0) = 1$ and the mixed
boundary conditions:
\begin{equation}  \label{eq:cond0}
-D \bigl(\partial_n S(r,\theta;t)\bigr)_{r=R} = 
\left\{ \begin{array}{l l} \kappa S(R,\theta;t) & (0\leq \theta < \ve) ,\\
0 &  (\ve \leq \theta \leq \pi) , \\ \end{array}  \right.
\end{equation}
where $\partial_n = \partial_r$ is the normal derivative directed
outwards the ball, $\Delta$ the Laplace operator, $D$ the diffusion
coefficient, $\kappa$ the intrinsic reactivity, and $\ve$ the angular
size of the escape window (or target site) at the North pole
(Fig. \ref{fig:1}).

The Laplace transform of the survival probability,
\begin{equation}
\tilde{S}(r,\theta; p) = \int\limits_0^\infty dt \, e^{-pt} \, S(r,\theta;t),
\end{equation}
satisfies the modified Helmholtz equation
\begin{equation}  \label{eq:Helm}
(p - D \Delta) \tilde{S}(r,\theta;p) = 1 ,
\end{equation}
subject to the same mixed boundary condition:
\begin{equation}  \label{eq:cond}
-D \bigl(\partial_r \tilde{S}(r,\theta;p)\bigr)_{r=R} = 
\left\{ \begin{array}{l l} \kappa \tilde{S}(R,\theta;p) &  (0\leq \theta < \ve) ,\\
0 &  (\ve \leq \theta \leq \pi) . \\ \end{array}  \right.
\end{equation}
According to the self-consistent approximation
\cite{Shoup81,Grebenkov17,Grebenkov18}, we substitute this mixed
boundary condition by an effective inhomogeneous Neumann condition:
\begin{equation}  \label{eq:condM}
-D \bigl(\partial_r \tilde{S}(r,\theta;p)\bigr)_{r=R} = Q \, \Theta(\ve - \theta),
\end{equation}
where $Q$ is an effective flux to be computed, and $\Theta$ is the
Heaviside step function.

We search the solution of the modified problem in \eqref{eq:Helm},
\eqref{eq:condM} in the form
\begin{equation}  \label{eq:ansatz}
\tilde{S}(r,\theta;p) = u_0(r) + \frac{R^2}{D} \sum\limits_{n=0}^\infty a_n g_n(r) P_n(\cos\theta) ,
\end{equation}
where $u_0(r)$ is the solution of \eqref{eq:Helm} with the Dirichlet
boundary condition, $u_0(r) = \frac{1}{p}(1 - g_0(r)/g_0(R))$,
$P_n(z)$ are the Legendre polynomials, $a_n$ are the unknown
coefficients to be determined, and $g_n(r)$ are the radial functions
satisfying
\begin{equation}
g''_n + \frac{2}{r} g'_n - \frac{n(n+1)}{r^2} g_n - \frac{p}{D} g_n = 0 ,
\end{equation}
{\clr where prime denotes the derivative with respect to $r$.}  A
solution of this equation, which is regular at $r = 0$, is the
modified spherical Bessel function of the first kind, which reads as
\begin{equation}  \label{eq:gn}
g_n(r) {\clr = i_n\bigl(r \sqrt{p/D}\bigr)} = \frac{\sqrt{\pi}}{\sqrt{2}} \, \frac{I_{n+1/2}\bigl(r \sqrt{p/D}\bigr)}{\bigl(r \sqrt{p/D}\bigr)^{1/2}} \,,
\end{equation}
where $I_\nu(z)$ is the modified Bessel function of the first kind,
and we fixed a particular normalisation (as shown below, the result
does not depend on the normalisation).

Substituting \eqref{eq:ansatz} into the modified boundary condition in
\eqref{eq:condM}, multiplying by $P_n(\cos\theta)\sin\theta$ and
integrating over $\theta$ from $0$ to $\pi$, one gets
\begin{equation}  \label{eq:Q}
2D\biggl(\frac{g'_0(R)}{p g_0(R)} - \frac{R^2}{D} a_0 g'_0(R)\biggr) = Q (1 - \cos\ve)   \quad (\textrm{for}~n = 0),
\end{equation}
and 
\begin{equation}
-2 R^2 a_n g'_n(R)  = Q \bigl(P_{n-1}(\cos\ve)-P_{n+1}(\cos\ve)\bigr)   \quad (\textrm{for}~ n > 0).
\end{equation}
Combining these results, we can express $a_n$ as
\begin{equation}  \label{eq:an}
a_n = - \frac{\phi_n(\ve) g'_0(R)}{g'_n(R)} \biggl(\frac{D}{pR^2} \frac{1}{g_0(R)} - a_0\biggr) \qquad (n = 1,2,\ldots),
\end{equation}
where
\begin{equation}  \label{eq:phi_def}
\phi_n(\ve) = \frac{P_{n-1}(\cos\ve) - P_{n+1}(\cos\ve)}{1 - \cos\ve}  \qquad (n = 1,2,\ldots).
\end{equation}

The remaining coefficient $a_0$ is determined by requiring that the
mixed boundary condition in \eqref{eq:cond} is satisfied {\it on
average} on the target, i.e.,
\begin{equation}
- D \int\limits_0^\ve d\theta \, \sin\theta \, \bigl(\partial_r \tilde{S}(r,\theta;p)\bigr)_{r=R} 
 = \int\limits_0^\ve d\theta \, \sin\theta \, \kappa \tilde{S}(R,\theta;p) .
\end{equation}
Using \eqref{eq:condM} and \eqref{eq:Q} for the left-hand side of this
relation and substituting \eqref{eq:ansatz} into the right-hand side,
one gets
\begin{equation}
\fl
2D\biggl(\frac{g'_0(R)}{p g_0(R)} - \frac{R^2}{D} a_0 g'_0(R)\biggr) = \kappa \frac{R^2}{D} \sum\limits_{n=0}^\infty a_n g_n(R) 
\frac{(1-\cos\ve)\phi_n(\ve)}{2n+1} \,,
\end{equation}
where we completed \eqref{eq:phi_def} by setting $\phi_0(\ve) \equiv
1$.  Expressing $a_n$ via \eqref{eq:an}, one obtains a closed
expression for $a_0$
\begin{equation}
a_0 = \frac{D}{p R^2 g_0(R)} \biggl(1 - \frac{g_0(R)}{Rg'_0(R)} \eta_p\biggr),
\end{equation}
where
\begin{equation}  \label{eq:eta}
\eta_p = \biggl(\frac{2D}{\kappa R(1-\cos\ve)} + \Rve_\ve \biggr)^{-1}\,,
\end{equation}
with
\begin{equation}
\Rve_\ve = \sum\limits_{n=0}^\infty \frac{g_n(R)}{R g'_n(R)} \, \frac{\phi_n^2(\ve)}{2n+1} \,.
\end{equation}
From \eqref{eq:an}, we get thus
\begin{equation}
a_n = - \frac{\phi_n(\ve)}{R g'_n(R)} \frac{D}{pR^2} \eta_p ,
\end{equation}
and finally
\begin{equation}
\tilde{S}(r,\theta;p) = \frac{1}{p} \biggl( 1 
- \eta_p  \sum\limits_{n=0}^\infty \phi_n(\ve) \frac{g_n(r)}{R g'_n(R)} P_n(\cos\theta) \biggr) .
\end{equation}

As a consequence, the Laplace-transformed probability density function
of the FPT, $\tilde{H}(r,\theta;p) = 1 - p \tilde{S}(r,\theta;p)$,
reads
\begin{equation}  \label{eq:Hp}
\tilde{H}(r,\theta;p) =  \eta_p \sum\limits_{n=0}^\infty \phi_n(\ve) \frac{g_n(r)}{R g'_n(R)} P_n(\cos\theta)  .
\end{equation}
As mentioned above, the solution does not depend on the normalisation
of functions $g_n(r)$, as they always enter as a ratio of $g'_n$ and
$g_n$.  One can check that the probability density function
$H(r,\theta;t)$ is correctly normalised, i.e.,
$\tilde{H}(r,\theta;p=0) = 1$.  In fact, one has in the limit $p\to
0$:
\begin{equation}
\frac{g_0(r)}{R g'_0(R)} \to \infty  , \qquad
\frac{g_n(r)}{R g'_n(R)} \to \frac{(r/R)^n}{n}   \quad (n=1,2,\ldots)
\end{equation}
so that the term with $n=0$ dominates both in $\eta_p$ and in the
series in \eqref{eq:Hp}.

When the starting point is uniformly distributed over a spherical
surface of radius $r$, the surface average yields
\begin{equation} \label{eq:Hpr}
\overline{ \tilde{H}(p)}_r \equiv \frac{1}{4\pi} \int\limits_0^\pi d\theta \, \sin \theta 
\int\limits_0^{2\pi} d\phi \,\tilde{H}(r,\theta;p)  
= \eta_p \frac{g_0(r)}{R g'_0(R)} \,.
\end{equation}
In turn, if the starting point is uniformly distributed in the volume
of the ball, one gets the volume average
\begin{eqnarray}  \nonumber
\overline{ \tilde{H}(p)} &\equiv& \frac{1}{4\pi R^3/3} \int\limits_0^R dr\, r^2
\int\limits_0^\pi d\theta \, \sin \theta \int\limits_0^{2\pi} d\phi \, \tilde{H}(r,\theta;p)  \\   \label{eq:Hpv}
&=& \frac{3\eta_p}{R \sqrt{p/D}}  \frac{g_1(R)}{R g_0'(R)} = \frac{3\eta_p}{R^2 p/D}\,. 
\end{eqnarray}

The explicit expressions \eqref{eq:Hp}, \eqref{eq:Hpr}, and
\eqref{eq:Hpv} for the Laplace-transformed PDF present one of the main
results.  We recall that \eqref{eq:Hp} is the exact solution of the
modified Helmholtz equation in \eqref{eq:Helm}, in which the mixed
boundary condition \eqref{eq:cond} is replaced by an effective
inhomogeneous Neumann condition \eqref{eq:condM}.  As a
consequence, \eqref{eq:Hp} and its volume and surface averages in
\eqref{eq:Hpr} and \eqref{eq:Hpv} are {\it approximate} solutions to
the original problem.  In \ref{sec:FEM}, we compare these approximate
solutions to the numerical solution of the original problem obtained
by a finite-elements method (FEM).  We show that the approximate
solutions are remarkably accurate for a broad range of $p$, for both
narrow ($\ve = 0.1$) and extended ($\ve = 1$) targets.

\begin{figure*}
\centering
\includegraphics[width=\textwidth]{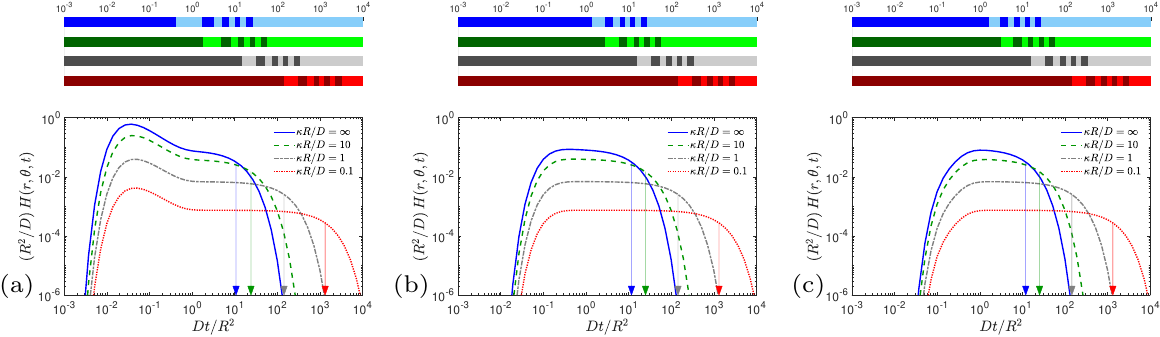}
\caption{
PDF $H(r,\theta;t)$ of the FRT, determined from \eqref{chief} to
\eqref{F} and rescaled by $R^2/D$, as function $D t/R^2$ for $r/R =
0.5$, $\ve = 0.1$, several values of dimensionless reactivity
$\bar{\kappa} = \kappa R/D$ (indicated in the plots), and $\theta = 0$
{\bf (a)}, $\theta = \pi/2$ {\bf (b)}, $\theta = \pi$ {\bf (c)}.
Self-consistent approximation is truncated to $n = 50$ terms.
Vertical arrows indicate the mean FRT, $t_{\rm mean}$.  Coloured
bar-codes above each figure indicate the cumulative depths
corresponding to four considered values of $\bar{\kappa}$, in
decreasing order from top to bottom.  Each bar-code is split into ten
regions of alternating brightness, each representing ten
$10\%$-quantiles of the distribution.  The numbers on top of the
bar-codes indicate the values of $D t/R^2$.  For example, in panel
{\bf (a)} the first dark blue region for the case $\bar{\kappa}
=\infty$ (perfect reaction) indicates that $10\%$ of reaction events
occur till $Dt/R^2 \simeq 3 \times 10^{-1}$.  The dimensionless mean
FRT in this case is around $10^1$, i.e., is almost two orders of
magnitude bigger that $Dt_{\rm mp}/R^2$, and over $70 \%$ of all
trajectories arrive to the target up to this time. }
\label{fig:fig2}
\end{figure*}

{\clr It is worth mentioning that the developed self-consistent
approximation can also be applied to the exterior problem when a
particle is released outside a ball of radius $R$ and searches for a
target on its boundary.  The derivation follows the same lines, with
two minor changes: (i) the modified Bessel function of the first kind
in Eq. \eqref{eq:gn} should be replaced by the modified Bessel
function of the second kind, i.e.
\begin{equation}  \label{eq:gn_ext}
g_n(r) = \frac{\sqrt{2}}{\sqrt{\pi}} \, \frac{K_{n+1/2}(r \sqrt{p/D})}{(r \sqrt{p/D})^{1/2}} \,,
\end{equation}
and (ii) the normal derivative is directed inside the sphere so that
the sign of all radial derivatives should be modified.  As a
consequence, the former expressions \eqref{eq:Hp}, \eqref{eq:Hpr},
\eqref{eq:Hpv} remain valid, except that 
Eq. \eqref{eq:eta} is replaced by
\begin{equation}  \label{eq:eta_ext}
\eta_p = \biggl(\Rve_\ve - \frac{2D}{\kappa R(1-\cos\ve)} \biggr)^{-1}\,.
\end{equation}
}

\section{Solution in time domain}
\label{apB}

To get the probability density function $H(r,\theta;t)$ in time
domain, we need to perform the inverse Laplace transform of
$\tilde{H}(r,\theta;p)$ from \eqref{eq:Hp}.  As diffusion occurs in a
bounded domain, the spectrum of the underlying Laplace operator is
discrete and is determined by the poles of $\tilde{H}(r,\theta;p)$,
considered as a function of a complex-valued parameter $p$.  We note
first that the zeros of the functions $g_0(R)$ and $g'_n(R)$, standing
in the denominator of $\tilde{H}(r,\theta;p)$, are not the poles, as
the corresponding divergence is compensated by vanishing of the
function $\eta_p$ at these points.  So, we are left with the poles of
the function $\eta_p$ from \eqref{eq:eta} which are determined by the
following equation on $p\in \C$:
\begin{equation}   \label{eq:poles_auxil}
\sum\limits_{n=0}^\infty \frac{\phi_n^2(\ve)}{2n+1} \, \frac{g_n(R)}{R g'_n(R)} = - \frac{2D}{\kappa R(1-\cos\ve)}  \,.
\end{equation}
As the eigenvalues are positive, the poles lie on the negative real
axis of the complex plane: $p < 0$.  Setting $R\sqrt{p/D} = i\alpha$,
one has
\begin{equation}
g_n(R) = i^n j_n(\alpha),  \qquad  R g'_n(R) = i^n \, \alpha j'_n(\alpha) ,
\end{equation}
where 
\begin{equation}
j_n(z) = \sqrt{\pi/2} \, \frac{J_{n+1/2}(z)}{\sqrt{z}}
\end{equation}
are the spherical Bessel functions of the first kind.

Substituting these relations into \eqref{eq:poles_auxil}, one gets the
following transcendental equation:
\begin{equation}  \label{eq:spectrum_eq}
F(\alpha) = - \frac{2D}{\kappa R(1-\cos\ve)}  \,,
\end{equation}
where
\begin{equation}  \label{eq:F_def}
F(\alpha) \equiv \sum\limits_{n=0}^\infty \frac{\phi_n^2(\ve)}{2n+1} \, \frac{j_n(\alpha)}{\alpha\, j'_n(\alpha)} \,.
\end{equation}
Let us denote by $\alpha_m$ all positive solutions of
\eqref{eq:spectrum_eq}.  The numerical computation of these zeros
employs the monotonous increase of $F(\alpha)$ that can be checked by
evaluating its derivative,
\begin{equation}
\fl
F'(\alpha) = \sum\limits_{n=0}^\infty \frac{\phi_n^2(\ve)}{2n+1} \, 
\frac{\alpha^2 [j'_n(\alpha)]^2 + \alpha j_n(\alpha) j'_n(\alpha) + [\alpha^2 - n(n+1)][j_n(\alpha)]^2}{\alpha^3 [j'_n(\alpha)]^2} ,
\end{equation}
and recognising that the numerator of the second factor is
proportional a positive integral:
\begin{equation}
\fl
\frac{2}{\alpha} \int\limits_0^\alpha dx \, x^2 \, j_n^2(x) = \alpha^2 [j'_n(\alpha)]^2 
+ \alpha j_n(\alpha) j'_n(\alpha) + [\alpha^2 - n(n+1)][j_n(\alpha)]^2 \,,
\end{equation}
which is actually the standard normalisation of spherical Bessel
functions.  For each $n = 0,1,2,\ldots$, let $\alpha_{nk}^0$ ($k =
0,1,2,\ldots$) denote all positive zeros of the function $j'_n(z)$.
Putting all these zeros in an increasing order, we call the elements
of this sequence by $\alpha_m^0$ (i.e., $\alpha_m^0 =
\alpha_{nk}^0$ for some indices $n$ and $k$).  At each $\alpha_m^0$,
the function $F(\alpha)$ diverges.  As a consequence, one can search
for zeros $\alpha_m$ on intervals between two successive zeros
$\alpha_m^0$ and $\alpha_{m+1}^0$.  As the function $F(\alpha)$ is
continuous on each such interval and monotonously increasing, there is
precisely one solution of \eqref{eq:spectrum_eq} on each interval.
The solution can be easily found by bisection or Newton's method.
This observation implies that all poles are simple.  Moreover, the
second eigenvalue $\lambda_1 = \alpha_1^2/R^2$ is bound from below as
\begin{equation}
\lambda_1 \geq (\alpha_{10}^0)^2/R^2 > 4/R^2,
\end{equation}
where $\alpha_1^0 = \alpha_{10}^0 = 2.081575978\ldots$ is the smallest
strictly positive zero of $j'_1(z)$ (i.e., the right-hand side is the
first strictly positive eigenvalue of the Laplace operator for the
full reflecting sphere).  As expected, this bound is independent of
the target size and determines the crossover time $t_c$ that we set as
\begin{equation}
t_c = \frac{R^2}{D} 
\end{equation}
(as a qualitative border between two regimes, $t_c$ is determined up
to an arbitrary numerical factor that we select here to be $1$).

The Laplace transform can now be inverted by the residue theorem that
yields
\begin{equation}  \label{eq:Ht_spectral}
H(r,\theta;t) = \frac{D}{R^2} \sum\limits_{m=0}^\infty  e^{-Dt\alpha_m^2/R^2}  \, v_m(r,\theta) ,
\end{equation}
where
\begin{equation}  \label{eq:vm}
v_m(r,\theta) \equiv \frac{2}{F'(\alpha_m)}  \sum\limits_{n=0}^\infty \phi_n(\ve) P_n(\cos\theta) \frac{j_n(\alpha_m r/R)}{-j'_n(\alpha_m)} \,.
\end{equation}
Note that the sign minus in front of $j'_n(\alpha_m)$ comes from the
change of variable:
\begin{equation}
\frac{\partial}{\partial p} F(\alpha) = F'(\alpha) \frac{d\alpha}{dp} = - F'(\alpha) \frac{R^2}{2D\alpha} \,.
\end{equation}

The explicit expression \eqref{eq:Ht_spectral} of the probability
density function is our main result.  From this spectral expansion,
one gets the survival probability
\begin{equation}  \label{eq:St_spectral}
S(r,\theta;t) = \sum\limits_{m=0}^\infty \frac{1}{\alpha_m^2} \, e^{-Dt\alpha_m^2/R^2} \, v_m(r,\theta) \,.
\end{equation}
One can also compute the volume- or surface-averaged PDFs, namely,
\begin{equation}  \label{eq:Ht_volume}
\overline{H(t)} = \frac{6D}{R^2} \sum\limits_{m=0}^\infty  \frac{e^{-Dt\alpha_m^2/R^2}}{\alpha_m F'(\alpha_m)} 
\end{equation}
and
\begin{equation}  \label{eq:Ht_surface}
\overline{H(t)}_r = \frac{2D}{R^2} \sum\limits_{m=0}^\infty 
e^{-Dt\alpha_m^2/R^2} \frac{j_0(\alpha_m r/R)}{-j'_0(\alpha_m)} \, \frac{1}{F'(\alpha_m)} \,.
\end{equation}
Another simpler expression is obtained for the starting point at the
centre of the sphere, in which case $v_m(0,0) = 2/F'(\alpha_m)$ and
thus
\begin{equation}
H(0,0;t) = \frac{D}{R^2} \sum\limits_{m=0}^\infty  \frac{e^{-Dt\alpha_m^2/R^2}}{F'(\alpha_m)}  \,.
\end{equation}

In the particular case $\ve = \pi$, the full sphere is reactive so
that $\phi_n(\pi) = \delta_{n,0}$, and \eqref{eq:F_def} is reduced to
$F(\alpha) = j_0(\alpha)/(\alpha j'_0(\alpha))$, while
\eqref{eq:spectrum_eq} is thus reduced to the Robin boundary
condition:
\begin{equation}  \label{eq:Robin_full}
\alpha j'_0(\alpha) + (\kappa R/D) j_0(\alpha) = 0 .
\end{equation}
In this case, \eqref{eq:vm} yields
\begin{equation}
v_m(r,\theta) = \frac{2\alpha_m \bar{\kappa}^2}{\alpha_m^2 + \bar{\kappa}^2 - \bar{\kappa}} \, \frac{j_0(\alpha_m r/R)}{j_1(\alpha_m)} \,,
\end{equation}
where we evaluated explicitly $F'(\alpha_m)$ by using
\eqref{eq:Robin_full}, and $\bar{\kappa} = \kappa R/D$.  As a consequence,
\eqref{eq:Ht_spectral} is reduced to the classical spectral
expansion of the probability density $H(r,\theta;t)$ of the FPT to a
partially reactive sphere.

\subsection{Partially reactive target}
\label{sec:asymp_partially}

When either the target is small ($\ve \ll 1$) or its reactivity is
small ($\kappa R/D \ll 1$), one can perform the perturbative analysis
to get approximate solutions of \eqref{eq:spectrum_eq}.  In fact, one
searches for a zero of this equation in the form
\begin{equation}
\alpha_{nk} = \alpha_{nk}^{0} + \delta \alpha_{nk}^{1} + O(\delta^2),
\end{equation}
where $\alpha_{nk}^{0}$ is the $k$-th zero of $j'_n(z)$, and
\begin{equation}
\delta = \frac{\kappa R(1-\cos\ve)}{2D}  \ll 1
\end{equation}
is the small parameter.  Expanding $j'_n(\alpha_{nk})$ and
$j_n(\alpha_{nk})$, using the properties of spherical Bessel functions
and identifying the leading terms, one gets $\alpha_{nk}^1$, from
which
\begin{equation}
\alpha_{nk} = \alpha_{nk}^0 \biggl(1 + \delta \frac{\phi_n^2(\ve)}{(2n+1)[(\alpha_{nk}^0)^2 - n(n+1)]} + O(\delta^2) \biggr) \,,
\end{equation}
which is valid for all $n$ and $k$ except $n = k = 0$.  The
computation for the case $n = k = 0$ is different because
$\alpha_{00}^0 = 0$, so that the last factor in the right-hand side is
undefined.  Since
\begin{equation}
\frac{j_0(z)}{z j'_0(z)} \simeq - \frac{3}{z^2} + O(1),
\end{equation}
one gets in this case
\begin{equation}
\alpha_0 = \alpha_{00} \simeq \sqrt{3\delta} + O(\delta) .
\end{equation}
Note that this correction is particularly important.  The leading
order of the corresponding eigenvalue reads as
\begin{equation}
\lambda_0 \simeq \frac{3\delta}{R^2} = \frac{3\kappa(1-\cos\ve)}{2RD} \,.
\end{equation}
As expected, when the target is small or its reactivity is small, the
Laplacian spectrum is close to that of the fully reflecting sphere.

\subsection{Perfectly reactive target}
\label{sec:asymp_perfectly}

When $\kappa = \infty$, the right-hand side of
\eqref{eq:spectrum_eq} is zero, and the asymptotic analysis for
small $\ve$ is different.  Let us first consider the smallest pole
which is expected to be small as $\ve \to 0$.  Since
\begin{equation}
\frac{j_n(\alpha)}{\alpha j'_n(\alpha)} \to \frac{1}{n}  \qquad (\alpha \to 0),
\end{equation}
one gets for small $\alpha$
\begin{equation}
- \frac{3}{\alpha^2} + \sum\limits_{n=1}^\infty \frac{\phi_n^2(\ve)}{n(2n+1)} = 0 .
\end{equation}
The asymptotic behaviour of the second term was given in
\cite{Grebenkov17}:
\begin{equation}
\sum\limits_{n=1}^\infty \frac{\phi_n^2(\ve)}{n(2n+1)} = \frac{32}{3\pi} \ve^{-1} + \ln(1/\ve) - \frac{7}{4} + \ln(2) + O(\ve) ,
\end{equation}
from which, in the leading order, one gets
\begin{equation}
\alpha_0^2 \simeq \frac{9\pi}{32} \ve \,.
\end{equation}
As discussed in \cite{Grebenkov17}, the numerical coefficient of this
asymptotic behaviour differs by around $10\%$ from the exact
coefficient $\pi/3$ \cite{Singer06a}.

\section{Asymptotic behaviour}

In this section, we discuss the long-time and the short-time
asymptotic behaviours of the probability density function
$H(r,\theta;t)$.

\subsection{Long-time behaviour}

As expected for diffusion in a bounded domain, the PDF exhibits an
exponential decay at long times, with the rate determined by the pole
with the smallest amplitude.  The latter is given by the smallest
positive solution of \eqref{eq:spectrum_eq} whose asymptotic behaviour
was discussed in \ref{sec:asymp_partially} and
\ref{sec:asymp_perfectly}.  In the narrow escape setting, this decay
rate is close to the inverse of the volume-averaged mean FRT
$\overline{t_{\rm mean}}$.  In turn, this mean, as well as
higher-order moments, can be obtained from $\tilde{H}(r,\theta;p)$ in
the limit $p\to 0$.

Using the basic properties of the functions $g_n$ (see
\ref{sec:basic}), we get
\begin{equation}
\Rve_\ve = \frac{3D}{pR^2} + \frac15 + \sum\limits_{n=1}^\infty \frac{\phi_n^2(\ve)}{n(2n+1)} + O(p) ,
\end{equation}
so that
\begin{equation} 
\fl
\eta_p = \frac{pR^2}{3D} \biggl\{ 1 - p \biggl(\frac{2R}{3\kappa(1-\cos\ve)} 
+ \frac{R^2}{3D} \biggl(\frac15 + 
\sum\limits_{n=1}^\infty \frac{\phi_n^2(\ve)}{n(2n+1)}\biggr) \biggr) + O(p^2)\biggr\}.
\end{equation}
As a consequence, the Laplace-transformed PDF behaves as
\begin{equation}
\tilde{H}(r,\theta;p) = 1 - p\, t_{\rm mean} + O(p^2),
\end{equation}
with the mean FRT 
\begin{equation}  
\fl
t_{\rm mean} = \frac{2R}{3\kappa(1-\cos\ve)} + \frac{R^2 - r^2}{6D} 
+ \frac{R^2}{3D} \sum\limits_{n=1}^\infty
\biggl(\frac{\phi_n^2(\ve)}{n(2n+1)} - \frac{\phi_n(\ve)}{n} (r/R)^n P_n(\cos\theta) \biggr).
\end{equation}
The first two terms provide the classical exact form of the mean FRT
when the target is the full sphere ($\ve = \pi$).  This result
generalises our earlier formula for the volume-averaged mean FRT
\cite{Grebenkov17}, which can be retrieved by averaging over the
starting point uniformly distributed in the bulk:
\begin{equation}  
\overline{t_{\rm mean}} = \frac{2R}{3\kappa(1-\cos\ve)} + \frac{R^2}{3D}\biggl(\frac15 + 
\sum\limits_{n=1}^\infty \frac{\phi_n^2(\ve)}{n(2n+1)} \biggr).
\end{equation}
In the narrow escape limit ($\ve \to 0$), this expression behaves as
\cite{Grebenkov17}
\begin{equation}  \label{m}
\overline{t_{\rm mean}} \simeq\frac{4R}{3\kappa}\ve^{-2} + \frac{32R^2}{9\pi D}\ve^{-1}+
\frac{R^2}{3D}\ln(1/\ve) + O(1).
\end{equation}
As mentioned earlier, the leading term in the latter expression is exact, while the numerical prefactor $32/(9\pi)$ in the subdominant term in this
approximate relation exceeds by $10\%$ the exact prefactor $\pi/3$
(for further details, see \cite{Grebenkov17}).

\subsection{Short-time behaviour}
\label{sec:Ashort}

\begin{figure}
\begin{center}
\includegraphics[width=0.49\textwidth]{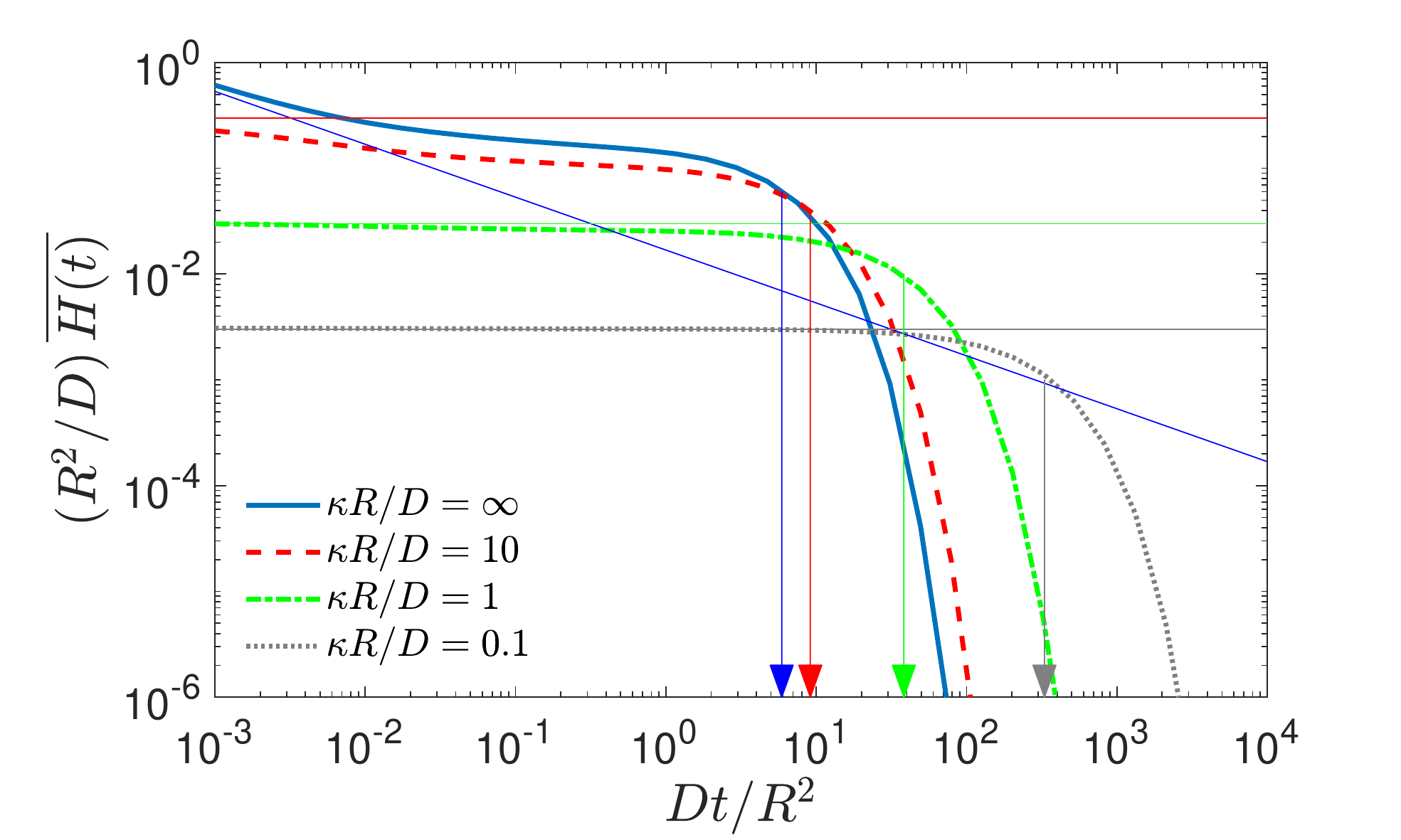}
\includegraphics[width=0.49\textwidth]{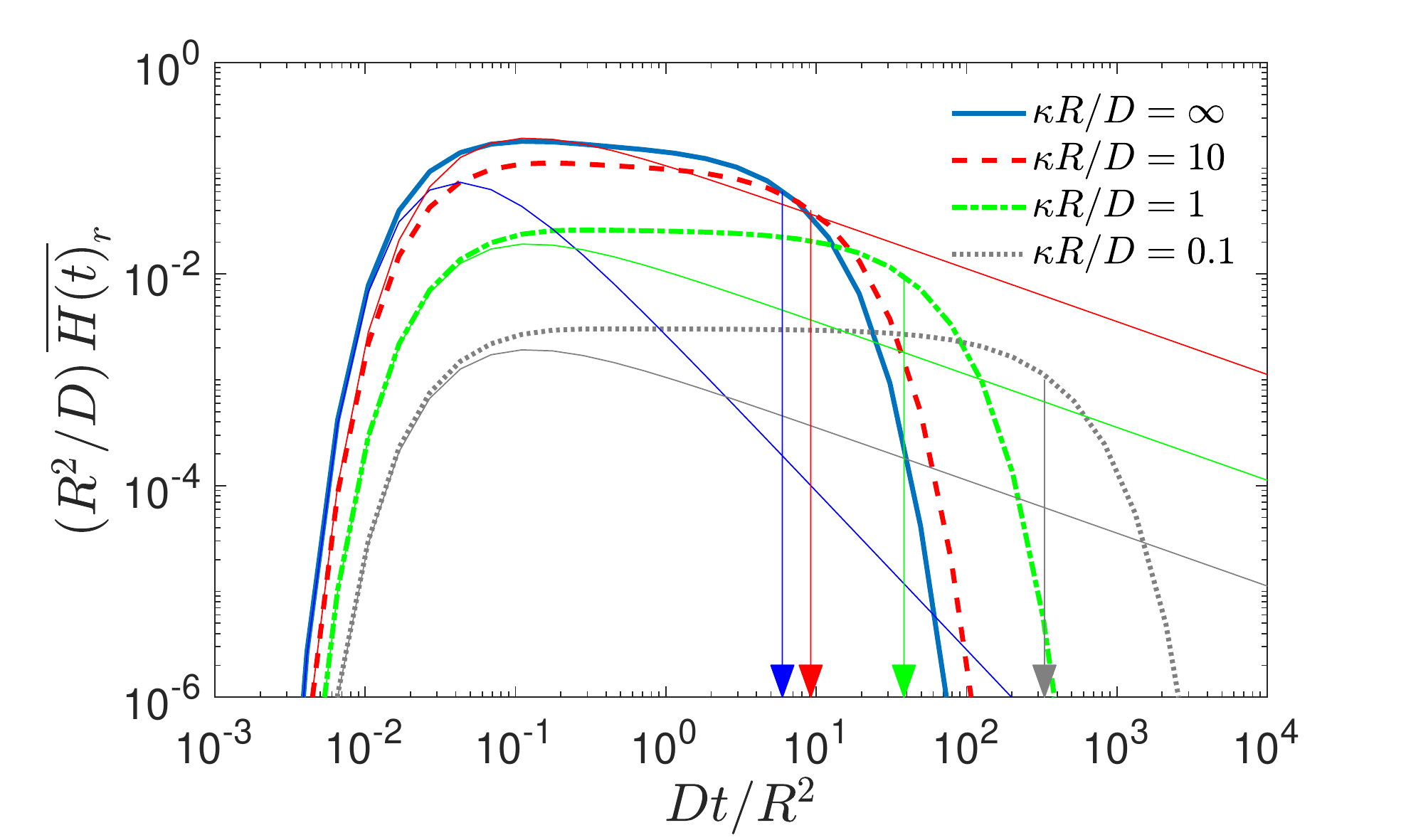}
\end{center}
\caption{
Volume-average (left) and surface-averaged (right) PDFs
$\overline{H(t)}$ and $\overline{H(t)}_r$ of the first reaction time
through a spherical cap, with $R = 1$, $r = 0.5$, $D = 1$, $\ve =
0.2$, and several values of $\kappa$.  Self-consistent approximation
is truncated to $n = 100$ terms.  Arrows indicate the mean FRT.  Thin
lines show the short-time asymptotic behaviours \eqref{eq:Hv_short}
and \eqref{eq:Hr_short} for two plots, respectively.}
\label{fig:Ht_r05}
\end{figure}

The short-time asymptotic behaviour is determined in the limit $p\to
\infty$.  We note that
\begin{equation}
\frac{g_0(R)}{R g'_0(R)} \simeq \frac{1}{\xi - 1} \simeq \frac{1}{\xi} \,,
\end{equation}
where $\xi = R\sqrt{p/D}$ (note that the first relation is very
accurate, up to exponentially small corrections).  Using
\eqref{eq:fn_inf}, one gets in the lowest order:
\begin{equation}  \label{eq:auxil11}
\Rve_\ve \simeq \frac{1}{\xi} + \sum\limits_{n=1}^\infty \frac{\phi_n^2(\ve)}{(2n+1) \bigl(\xi - 1 + \frac{n(n+1)}{2\xi}\bigr)} 
\simeq \frac{2}{\xi(1-\cos\ve)}  \,,
\end{equation}
where we used the identity
\begin{equation}  \label{eq:phin_ident}
\sum\limits_{n=1}^\infty \frac{\phi_n^2(\ve)}{2n+1} = \frac{1+\cos\ve}{1-\cos\ve} 
\end{equation}
(the next-order terms in \eqref{eq:auxil11} seems to be $\xi^{-2}$,
possibly with logarithmic corrections).  
We get thus
\begin{equation}  \label{eq:etap_large}
\eta_p \simeq \frac{R(1-\cos\ve)}{2(D/\kappa + \sqrt{D/p})}  \qquad (p\to\infty) \,,
\end{equation}
from which
\begin{equation}  \label{eq:Hv_short}
\overline{H(t)} \simeq \frac{3(1-\cos\ve)}{2R} \times \left\{ \begin{array}{l l}  \kappa & (\kappa < \infty) ,\\
\sqrt{D/(\pi t)} & (\kappa = \infty). \\ \end{array} \right.
\end{equation}
Figure \ref{fig:Ht_r05}(left) illustrates the accuracy of this
asymptotic behaviour.

Since 
\begin{equation}  \label{eq:g0rg0R_new}
\frac{g_n(r)}{Rg'_n(R)}  \simeq \frac{e^{-(R-r)\sqrt{p/D}}}{r \sqrt{p/D}} \qquad (p\to \infty) ,
\end{equation}
we find
\begin{equation}   \label{eq:tildeHr_short}
\overline{\tilde{H}(p)}_r \simeq \frac{R(1-\cos\ve)}{2r(1 + \sqrt{pD}/\kappa)} e^{-(R-r)\sqrt{p/D}} ,
\end{equation}
from which
\begin{equation}  \label{eq:Hr_short}
\fl
\overline{H(t)}_r \simeq \frac{R(1-\cos\ve)}{2r \sqrt{\pi D t}} e^{-(R-r)^2/(4Dt)} \times 
\left\{ \begin{array}{l l} \kappa & (\kappa < \infty) ,\\
(R-r)/(2t) & (\kappa = \infty) . \\ \end{array} \right.
\end{equation}
Figure \ref{fig:Ht_r05}(right) illustrates the accuracy of this
asymptotic behaviour.

In the particular case $r = 0$, one gets
\begin{equation} 
\overline{\tilde{H}(p)}_{r=0} \simeq R(1-\cos\ve) e^{-R\sqrt{p/D}} \times 
\left\{ \begin{array}{l l} \kappa/D & (\kappa < \infty) ,\\
\sqrt{p/D} & (\kappa = \infty) , \\  \end{array} \right.
\end{equation}
from which
\begin{equation}  \label{eq:Hr_short0}
\overline{H(t)}_{r=0} \simeq \frac{R(1-\cos\ve)}{\sqrt{4\pi D t^3}} e^{-R^2/(4Dt)} 
\times \left\{ \begin{array}{l l} \kappa R/D & (\kappa < \infty) ,\\
\frac{R^2}{2Dt} - 1 & (\kappa = \infty).  \\ \end{array} \right.
\end{equation}

Finally, the short-time asymptotic behaviour of $H(r,\theta;t)$ is
obtained by combining \eqref{eq:etap_large} and
\eqref{eq:g0rg0R_new} to get
\begin{equation*}
\tilde{H}(r,\theta;p) \simeq \frac{R(1-\cos\ve)}{2r(1 + \sqrt{pD}/\kappa)} e^{-(R-r)\sqrt{p/D}}  
 \biggl(1 + \sum\limits_{n=1}^\infty \phi_n(\ve) \, P_n(\cos\theta) \biggr) .
\end{equation*}
The sum in the parentheses can be computed exactly by noting that
\begin{equation*}
\fl
P_{n-1}(x) - P_{n+1}(x) = \frac{2n+1}{n(n+1)} (1-x^2) P'_n(z) 
= (2n+1) \int\limits_x^1 dx' P_n(x')
\end{equation*}
and using the completeness of the Legendre polynomials:
\begin{equation}
\sum\limits_{n=0}^\infty \frac{2n+1}{2}\, P_n(x)\, P_n(y) = \delta(x-y) .
\end{equation}
We get thus
\begin{equation}  \label{eq:Hp_short2}
\tilde{H}(r,\theta;p) \simeq \frac{R e^{-(R-r)\sqrt{p/D}}}{r(1 + \sqrt{pD}/\kappa)} \,  \Theta(\ve - \theta)  \qquad (p\to \infty),
\end{equation}
where $\Theta(x)$ is the Heaviside function.  The inverse Laplace
transform reads
\begin{equation}  \label{eq:Ht_short2}
\fl
H(r,\theta;t) \simeq \frac{R \, \Theta(\ve - \theta)}{r \sqrt{\pi D t}} e^{-(R-r)^2/(4Dt)} \times 
\left\{  \begin{array}{l l}  \kappa & (\kappa < \infty) ,\\
(R-r)/(2t) & (\kappa = \infty).  \\  \end{array} \right.
\end{equation}
The Heaviside function distinguishes two cases.  When the starting
point lies within the solid angle of the target (i.e., $\theta <
\ve$), the distance to the target is just $R-r$, and this asymptotic
formula holds.  In turn, when the starting point lies outside this
solid angle (i.e., $\theta > \ve$), the distance to the target is
larger than $R-r$, $\tilde{H}(r,\theta;p)$ should decay faster, the
contributions in \eqref{eq:Hp_short2} and \eqref{eq:Ht_short2} are
thus cancelled, and one needs more refined asymptotic analysis.
Without dwelling further on this analysis, we present below
probabilistic arguments to get a reasonable approximation.

When the starting point is far from the small escape window,
small exit times are extremely unlikely, with the probability density
vanishing exponentially fast as $\exp(-\sigma^2/(4Dt))$, where $\sigma
\gg \sqrt{Dt}$ is the distance to the escape window.  To find the
correct prefactor to this exponential tail, we note that (i) a small
spherical cap, seen from far away, can approximated by a small sphere
of the same radius $\rho = 2R\sin(\ve/2)$, centred at the North pole;
the distance to the boundary of this sphere is $\sigma = \sqrt{R^2 +
r^2 - 2rR \cos\theta} - \rho$; and (ii) the reflecting boundary of the
spherical confining domain can be removed because only almost straight
trajectories to the target, not touching the confining boundary, do
matter in the short-time limit.  In other words, the short-time
behaviour of the density $H(r,\theta; t)$ should be close to that of
the density $H_\infty(\sigma;t)$ for diffusion in the exterior of a
small spherical target.  The latter was derived by Collins and Kimball
\cite{Collins49},
\begin{eqnarray}   \nonumber
H_\infty(\sigma;t) &=&\frac{\kappa}{\sigma+\rho}\exp\biggl(-\frac{\sigma^2}{4Dt}\biggr)
\biggl\{\frac{\rho}{\sqrt{\pi Dt}}  \\   \label{eq:Ht_Rinf}
&-&\biggl(1+\frac{\kappa\rho}{D}\biggr)\erfcx\biggl(\frac{\sigma}{\sqrt{4Dt}}+\biggl(1+\frac{\kappa \rho}{D}\biggr)
\frac{\sqrt{Dt}}{\rho}\biggr)\biggr\},
\end{eqnarray}
where $\erfcx(x)=e^{x^2} \erfc(x)$ is the scaled complementary error
function (see also discussion in \cite{Grebenkov18b}).  In the limit
$\kappa\to\infty$, this expression reduces to
\begin{equation}
\label{eq:Ht_Rinf_kinf}
H_\infty(\sigma;t)=\frac{\rho}{\sigma+\rho} \, \frac{\sigma}{\sqrt{4\pi Dt^3}} \, \exp\biggl(-\frac{\sigma^2}{4Dt}\biggr) .
\end{equation}
In the limit $\sqrt{Dt} \gg \max\{\sigma,\rho\}$,
\eqref{eq:Ht_Rinf} yields  the long-time behaviour in
\eqref{eq:Levy-Smirnov}.

\begin{figure}
\begin{center}
\includegraphics[width=\textwidth]{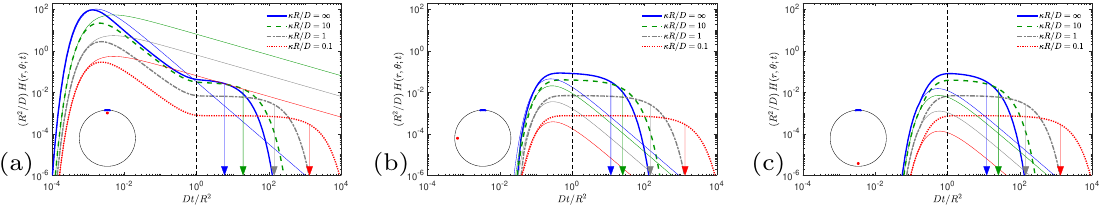}
\end{center}
\caption{
PDF $H(r,\theta; t)$, rescaled by $R^2/D$, as shown on
Fig. \ref{fig:fig22}, is compared to its short-time approximation
\eqref{eq:Ht_short2} for $\theta= 0$ {\bf (a)} and \eqref{eq:Ht_Rinf}
for $\theta = \pi/2$ {\bf (b)} and $\theta = \pi$ {\bf (c)}, shown by
thin lines of the same colour.  The other parameters are kept:
$r/R=0.9$, $\ve=0.1$, $\bar{\kappa}=\kappa R/D$ as indicated in the panels.
Vertical arrows indicate the rescaled mean FRT $Dt_{\rm mean}/R^2$,
while the dashed vertical line shows the rescaled crossover time
$Dt_c/R^2 = 1$.}
\label{fig:Ht_r09_eps01_short}
\end{figure}

Figure \ref{fig:Ht_r09_eps01_short} reproduces the rescaled PDF from
Fig. \ref{fig:fig22} to illustrate the quality of the short-time
asymptotic formulae.  For the case $\theta = 0$, the panel (a)
confirms that the short-time asymptotic formula \eqref{eq:Ht_short2}
accurately captures the left exponential tail of the PDF up to the
most probable time, i.e., for $t \lesssim t_{\rm mp}$.  However, this
approximation fails at longer times due to the confinement.  Note also
that the estimation of the most probable time $t_{\rm mp}$ from
\eqref{eq:Ht_short2}, suggested in \cite{Grebenkov18b} for a simpler
geometric setting, is not accurate here, except for the perfectly
reactive target.  From Fig. \ref{fig:Ht_r09_eps01_short}(a), we only can
say that $t_{\rm mp} < (R-r)^2/(6D)$ for the perfectly reactive target
($\kappa = \infty$), and $t_{\rm mp} < (R-r)^2/(2D)$ for partially
reactive targets ($\kappa < \infty$).
The panels (b) and (c) illustrate the accuracy of another short-time
asymptotic formula, given by \eqref{eq:Ht_Rinf}, which is applicable
when the starting point is far away from the target ($\theta = \pi/2$
and $\theta = \pi$, respectively).  Once again, the left exponential
tail of $H(r,\theta; t)$ is captured reasonably well.  In this case,
there is no hump, and a plateau region emerges immediately after this
tail.  As the most probable time now belongs to this plateau region,
it is very difficult to quantify.  This is a radically new feature of
the narrow escape problem, as compared to recent works
\cite{Grebenkov18,Grebenkov18b}.

\subsection{Explicit closed-form approximation}
\label{sec:approximation}

The knowledge of both short-time and long-time asymptotic behaviours
of the probability density motivated the search of closed-form
approximations for the probability density $H(r,\theta; t)$ over the
whole range of times.  For instance, Isaacson and Newby developed a
uniform asymptotic approximation for perfectly reactive small targets
in bounded domains using a short-time correction calculated by a
pseudopotential approximation \cite{Isaacson13}.  In turn, a linear
superposition of the short-time approximation in
\eqref{eq:Ht_Rinf} with the long-time exponential decay was
proposed in \cite{Grebenkov18b} for a partially reactive spherical
target located in the centre of a larger confining sphere.

Here, we rationalise and improve the linear superposition by employing
the simple idea of splitting trajectories toward a small target into
two groups: (i) ``direct'' trajectories that do not hit the boundary
of the confining domain and thus do not ``know'' about its presence,
and (ii) ``indirect'' trajectories that reach the boundary and thus
explore the {\it bounded} confining domain until eventual reaction
with the target.  The statistics of direct trajectories is close to
that of trajectories in the unbounded exterior of the target; in
particular, if the starting point is far from a small target, the
probability density of the first exit (reaction) times is close to
$H_\infty(\sigma;t)$ given by \eqref{eq:Ht_Rinf}.  In turn, the
indirect trajectories are responsible for the long-time exponential
decay of the probability density.  If $\tau_1$ and $\tau_2$ denote the
{\it conditional} first exit times for such direct and indirect
trajectories respectively, then one can set $\tau = \tau_1$ with the
probability $q_1$ of realising a direct trajectory, and $\tau =
\tau_2$ with the probability $q_2 = 1 - q_1$ of realising an indirect
trajectory.  If the related probability densities $\rho_1(t)$ and
$\rho_2(t)$ were known, then the probability density of $\tau$ would
be simply
\begin{equation}  \label{eq:rhot_approx}
H(r,\theta; t) = q_1 \rho_1(t) + q_2 \rho_2(t) .
\end{equation}

\begin{figure}
\begin{center}
\includegraphics[width=\textwidth]{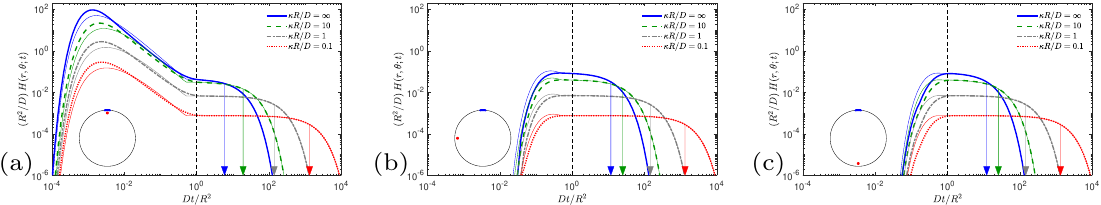}
\end{center}
\caption{
PDF $H(r,\theta; t)$, rescaled by $R^2/D$, as shown on
Fig. \ref{fig:fig22}, is compared to the approximation in
\eqref{eq:rhot_approx} for $\theta= 0$ {\bf (a)}, $\theta = \pi/2$
{\bf (b)} and $\theta = \pi$ {\bf (c)}, shown by thin lines of the
same colour.  The other parameters are kept: $r/R=0.9$, $\ve=0.1$,
$\bar{\kappa}=\kappa R/D$ as indicated in the panels.  Vertical arrows
indicate the mean FRT $t_{\rm mean}$, while the dashed vertical line
shows the crossover time $t_c = R^2/D$.  We set $T = 0.5 t_c$.}
\label{fig:Ht_r09_eps01_app}
\end{figure}

The computation of the probability densities $\rho_1(t)$ and
$\rho_2(t)$ is an even more complicated problem than the original
mixed boundary problem for $H(r,\theta; t)$.  So, we propose another,
much simpler splitting into two groups by introducing a heuristic
threshold $T$ for the first exit time.  In this approximation, all
trajectories toward the target that are shorter than $T$ are treated
as ``direct'', whereas all longer trajectories are treated as
``indirect''.  According to this construction, the first exit times of
direct trajectories, that can be described by the probability density
$H_\infty(\sigma;t)$, are conditioned to be shorter than $T$; in turn,
the first exit times of indirect trajectories, that can be described
by the probability density $\mu e^{-\mu t}$ with $\mu = D\lambda_0$
being related to the smallest eigenvalue $\lambda_0$ of the Laplace
operator in the confining domain, are conditioned to be longer than
$T$.  The related conditional probability densities are then given as
\begin{subequations}  \label{eq:rho1_rho2}
\begin{eqnarray} 
\rho_1(t) &=& \Theta(T-t) H_\infty(\sigma;t)/(1-S_\infty(\sigma;T)) , \\
\rho_2(t) &=& \Theta(t-T) \mu e^{-\mu (t-T)}, 
\end{eqnarray}
\end{subequations}
where 
\begin{eqnarray}  \nonumber
S_\infty(\sigma;t) &=& 1 - \frac{\rho
\exp\bigl(-\frac{\sigma^2}{4Dt}\bigr)}{(\sigma+\rho)(1 + D/(\kappa\rho))}
\biggl\{ \erfcx\biggl(\frac{\sigma}{\sqrt{4Dt}}\biggr)  \\    \label{eq:S_inf}
&-& \erfcx\biggl(\frac{\sigma}{\sqrt{4Dt}} + \left(1 + \frac{\kappa \rho}{D}\right) \frac{\sqrt{Dt}}{\rho}\biggr)\biggr\} 
\end{eqnarray}
is the survival probability obtained by integrating
$H_\infty(\sigma;t)$.  To ensure the continuity of $\rho(t)$ in
\eqref{eq:rhot_approx}, we set
\begin{equation}
q_1 = \frac{\mu e^{-\mu T} (1 - S_\infty(\sigma;T))}{H_\infty(\sigma;T) e^{-\mu T} + \mu e^{-\mu T} (1 - S_\infty(\sigma;T))} \,.
\end{equation}
The threshold $T$ can either be fixed to a prescribed value (e.g., $T
= t_c$), or determined by imposing an additional condition, i.e., the
continuity of the derivative of $\rho(t)$.  Figure
\ref{fig:Ht_r09_eps01_app} illustrates the accuracy of this
{\it explicit} approximation.

\section{Basic properties and implementation}
\label{sec:basic}

The functions $g_n$ and their derivatives can be computed via the
recurrent relations
\begin{eqnarray}
g_0(r) &=&  \frac{\sinh \xi}{\xi} \,,   \qquad g_1(r) = \frac{\xi \cosh \xi - \sinh \xi}{\xi^2} \,, \\
g_{n+1}(r) &=& g_{n-1}(r) - \frac{2n+1}{\xi} g_n(r) \quad (n = 1,2,\ldots),  \\
g'_0(r) &=& \sqrt{p/D} \, g_1(r),  \\
g'_n(r) &=& \sqrt{p/D} \biggl(g_{n-1}(r) - \frac{n+1}{\xi} g_n(r)\biggr) ,
\end{eqnarray}
with $\xi = r \sqrt{p/D}$.  As a consequence, one has
\begin{eqnarray}
r \frac{g'_0(r)}{g_0(r)} &=& r\sqrt{p/D} \, \frac{g_1(r)}{g_0(r)} =  \frac{\xi^2}{f_1}  \,, \\  \label{eq:ggn_fn}
r \frac{g'_n(r)}{g_n(r)} &=& f_n - (n+1) \quad (n=1,2,\ldots) ,
\end{eqnarray}
where new functions
\begin{equation}
f_n = \xi \, \frac{g_{n-1}(r)}{g_n(r)}
\end{equation}
satisfy the recurrent relations
\begin{eqnarray}
f_1 &=& \frac{\xi^2}{\xi \coth \xi - 1} \,, \\   \label{eq:fn_recus}
f_n &=& \frac{\xi^2}{f_{n-1} - (2n-1)}  \quad (n=2,3,\ldots) .
\end{eqnarray}
Unfortunately, these relations are numerically unstable for $\xi$
below $n$ that prohibits their use even for moderate $n$.  

In order to resolve this numerical problem, we propose the following
trick.  Rewriting \eqref{eq:fn_recus} as
\begin{equation}  \label{eq:fn_recur2}
f_n = 2n+1 + \frac{\xi^2}{f_{n+1}} \,, 
\end{equation}
one can consider the recurrent relations for $f_n$ as a continued
fraction representation:
\begin{equation}
\xi \coth \xi = 1 + \frac{\xi^2}{\displaystyle 3 + \frac{\xi^2}{\displaystyle 5 + \frac{\xi^2}{7 + \cdots}}} \,.
\end{equation}
To overcome numerical instabilities, we propose the following
algorithm: (i) for any $\xi$, we select a large enough truncation
parameter $n_{\rm max}$ such that $n_{\rm max} \gg |\xi|$ (in
practice, we set $n_{\rm max} = \max\{ 100, [10\xi]\}$); (ii) we
approximate $f_n$ with $n = n_{\rm max}$ according to its Taylor
expansion
\begin{eqnarray}  \label{eq:fn_Taylor}
\fl
f_n &=& 2n+1 + \frac{\xi^2}{2n+3} - \frac{\xi^4}{(2n+3)^2(2n+5)} 
+ \frac{2\xi^6}{(2n+3)^3 (2n+5)(2n+7)} \\  \nonumber
\fl
&-& \frac{(10n+27)\xi^8}{(2n+3)^4 (2n+5)^2(2n+7)(2n+9)} + O(\xi^{10}),
\end{eqnarray}
which is valid whenever $|\xi| \ll n$ (even if $|\xi|$ is large);
(iii) applying \eqref{eq:fn_recur2} in a backward direction, we
compute all $f_n$ with $1 \leq n \leq n_{\rm max}$.  When the argument
$|\xi|$ is very large (as compared to $n$), one can use the asymptotic
relation
\begin{equation}  \label{eq:fn_inf}
f_n = \xi + n + \frac{n(n+1)}{2\xi} + O(\xi^{-2}) \qquad (\xi\to\infty).
\end{equation}

We checked that this algorithm provides very accurate results for real
$\xi$.  Qualitatively, the improved numerical stability of recurrent
relations in the backward direction resembles the advantage of an
implicit Euler scheme for solving differential equations over an
explicit one.  Note that the computation can be more subtle for some
imaginary $\xi$, at which $\coth \xi$ diverges (i.e., at $\xi = i\pi
k$ with an integer $k$).  
However, we have not experienced such numerical difficulties for
considered parameters.

We note that \eqref{eq:ggn_fn} and \eqref{eq:fn_Taylor} imply
\begin{equation}
r \frac{g'_n(r)}{g_n(r)} = n + \frac{\xi^2}{2n+3} - \frac{\xi^4}{(2n+3)^2(2n+5)} + O(\xi^6),
\end{equation}
which could also be obtained directly from 
\begin{equation}
g_n(r) = \frac{\sqrt{\pi}}{2} \sum\limits_{k=0}^\infty \frac{(z/2)^{n+2k}}{k! \, \Gamma(n+k+1/2)} 
\end{equation}
(with $z = r \sqrt{p/D}$).  One sees that for $z \geq 0$, all
functions $g_n(r)$ are positive and monotonously increasing because
\begin{equation}
g'_n(r) = \sqrt{p/D}\biggl( \frac{n}{2n+1} g_{n-1}(r) + \frac{n+1}{2n+1} g_{n+1}(r)\biggr) \geq 0.
\end{equation}

We also define $\hat{f}_n(\xi) = f_n(i\xi)$ so that
\begin{equation}
\frac{Rg'_0(R)}{g_0(R)} = - \frac{\alpha^2}{\hat{f}_1} \,,  \quad
\frac{Rg'_n(R)}{g_n(R)} = \hat{f}_n - (n+1) ,
\end{equation}
with $\hat{f}_n(\alpha) = \alpha \, j_{n-1}(\alpha)/j_n(\alpha)$ for
$n = 1,2,3,\ldots$, which yields
\begin{equation}
\hat{f}_1 = \frac{\alpha^2}{1 - \alpha\, \ctan \alpha} \,,  \qquad 
\hat{f}_n = \frac{\alpha^2}{2n-1 - \hat{f}_{n-1}} \,.
\end{equation}
In particular, one gets
\begin{equation}
F(\alpha) = - \frac{\hat{f}_1(\alpha)}{\alpha^2} + \sum\limits_{n=1}^\infty \frac{\phi_n^2(\ve)}{2n+1} \, \frac{1}{\hat{f}_n(\alpha) - (n+1)}   \,.
\end{equation}

Note that these recurrent relations yield very accurate results for
$\alpha \gtrsim n$.  For smaller $\alpha$, one can use the Taylor
expansion of $\hat{f}_n$.

\section{Numerical verification}
\label{sec:FEM}

\begin{figure}
\begin{center}
\includegraphics[width=0.49\textwidth]{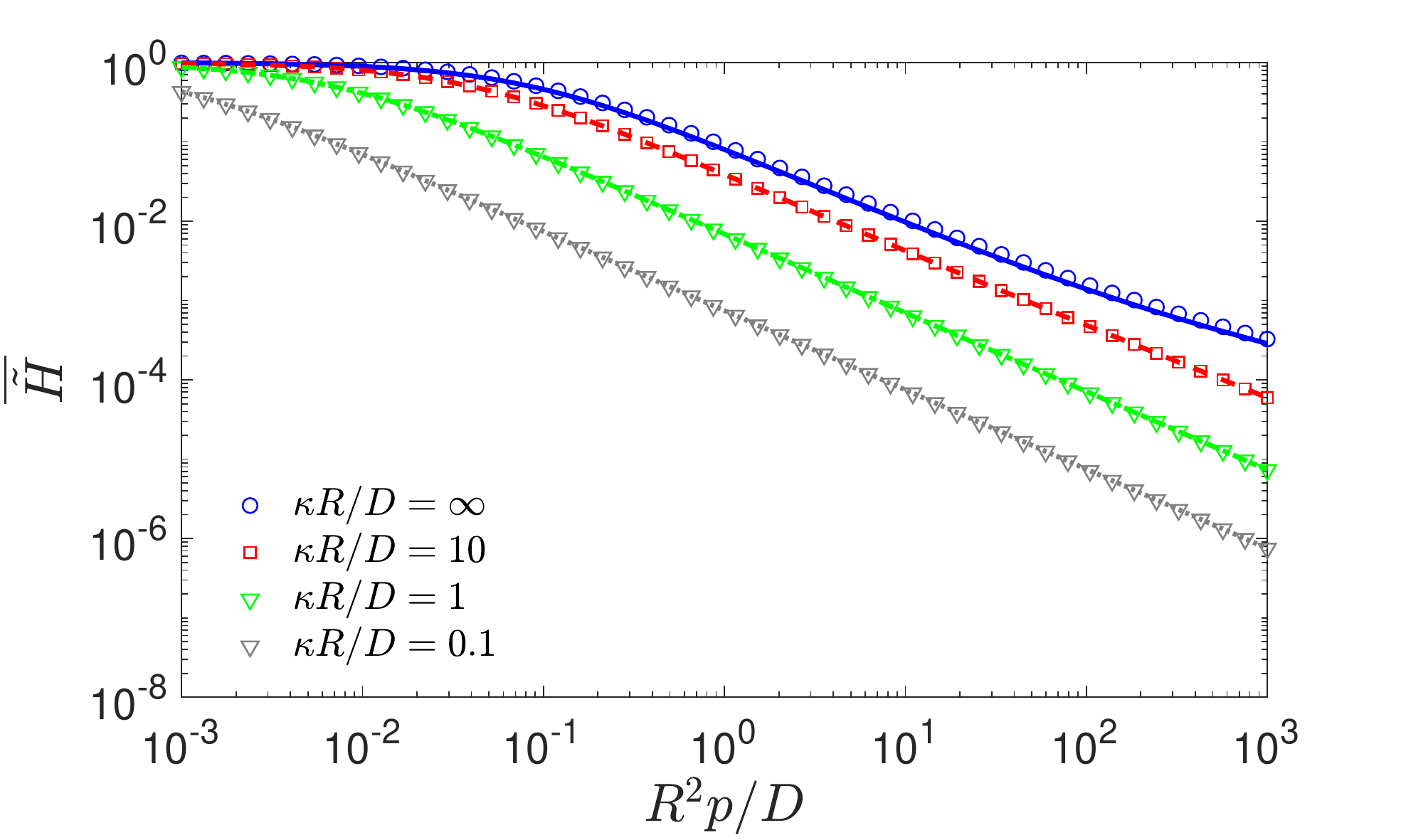}
\includegraphics[width=0.49\textwidth]{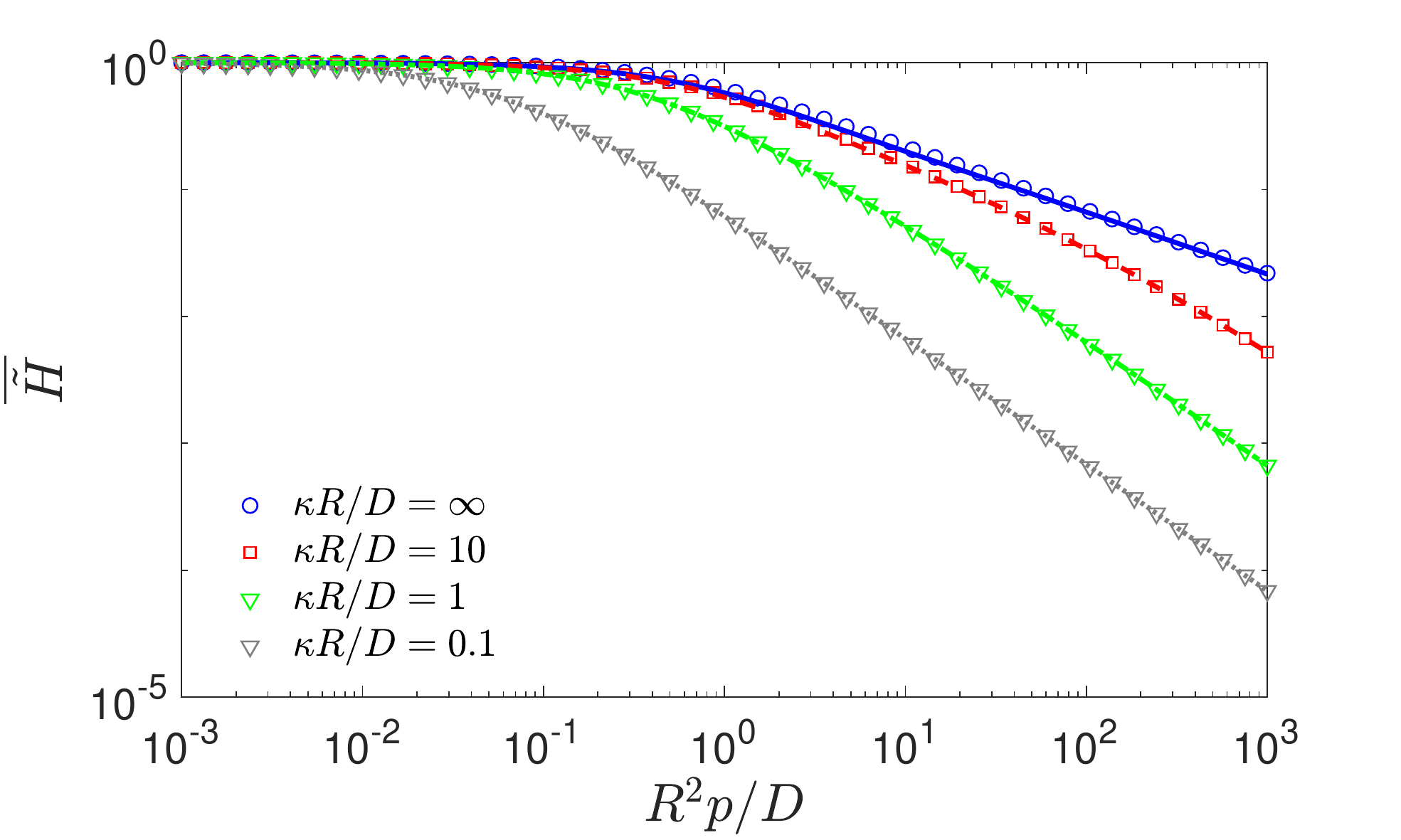}
\end{center}
\caption{
Volume-averaged Laplace-transformed PDF $\overline{\tilde{H}(p)}$ of
the FRT through a spherical cap, with $R = 1$, $D = 1$, $\ve = 0.1$
(left) and $\ve = 1$ (right), and several values of $\kappa$.
Self-consistent approximation (lines), truncated to $n = 1000$ terms,
is compared to the FEM solution of the original problem (symbols),
with $h_{\rm max} = 0.01$. Note that our theoretical predictions are
almost indistinguishable from the FEM solution for several decades of
variation of the dimensionless parameter $R^2 p/D$.}
\label{fig:FEM}
\end{figure}

\begin{figure}
\begin{center}
\includegraphics[width=0.49\textwidth]{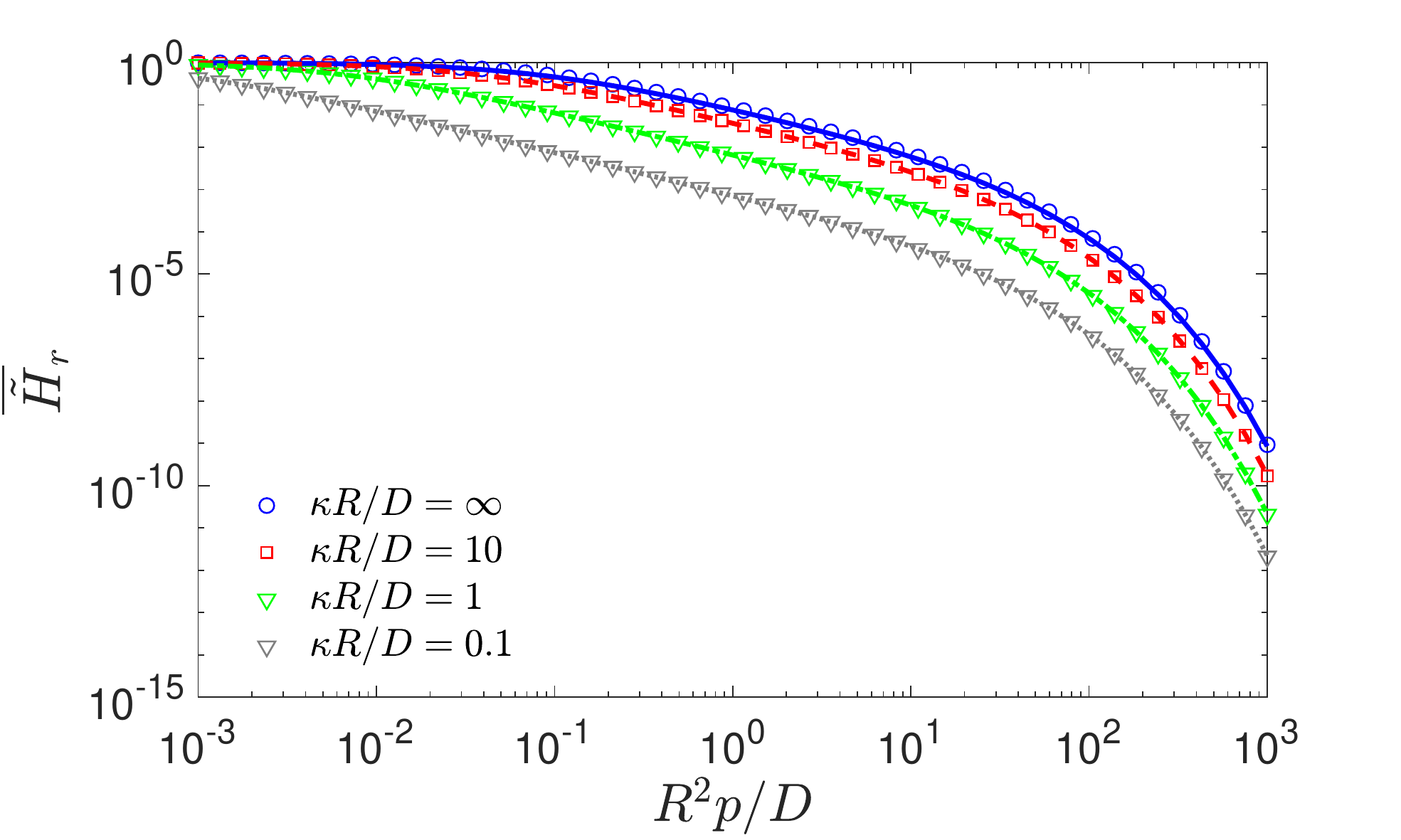}
\includegraphics[width=0.49\textwidth]{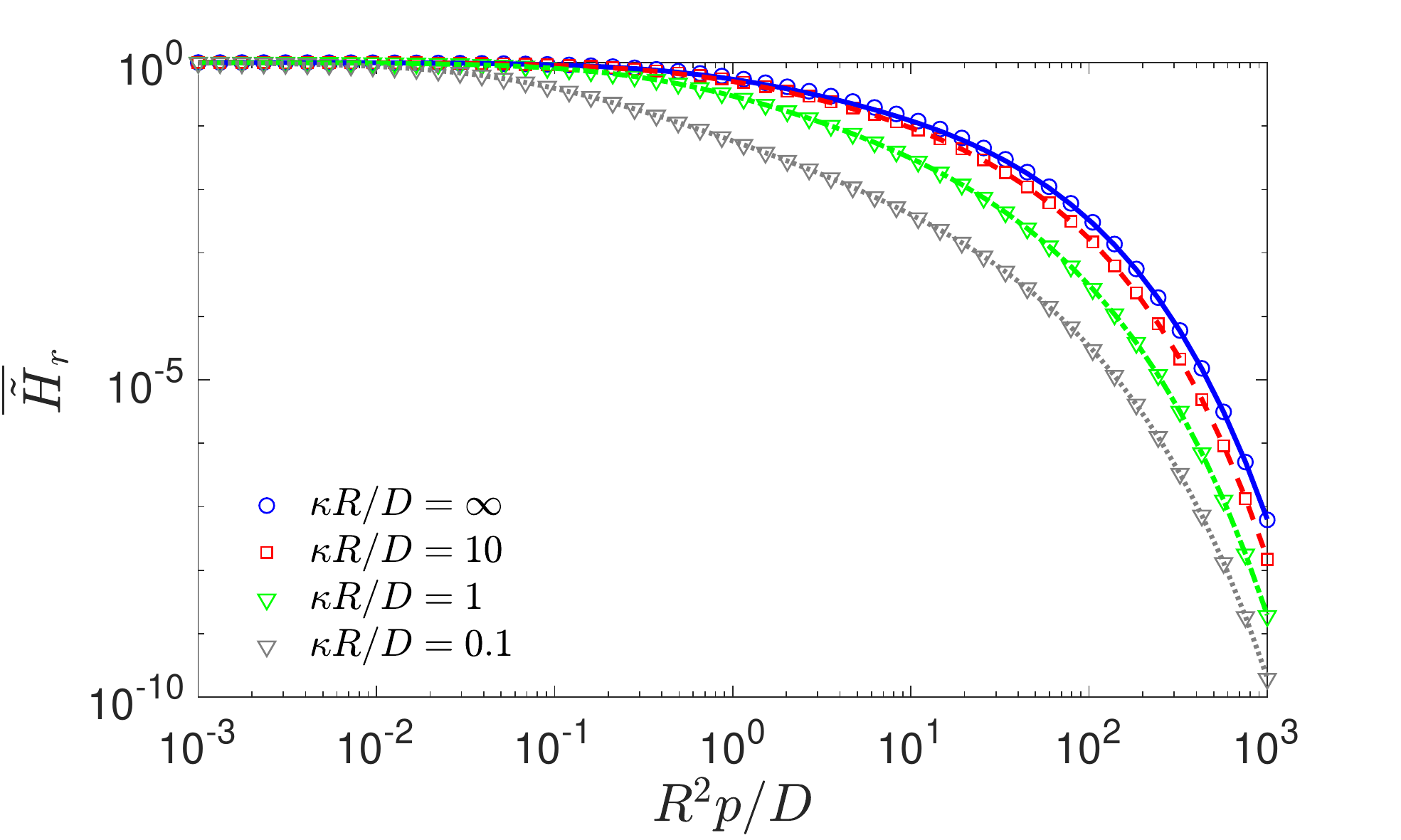}
\end{center}
\caption{
Surface-averaged Laplace-transformed PDF $\overline{\tilde{H}(p)}_r$
of the FRT through a spherical cap, with $R = 1$, $D = 1$, $r =
0.5$, $\ve = 0.1$ (left) and $\ve = 1$ (right), and several values of
$\kappa$.  Self-consistent approximation (lines), truncated to $n =
1000$ terms, is compared to the FEM solution of the original problem
(symbols), with $h_{\rm max} = 0.01$. Note that our theoretical
predictions are almost indistinguishable from the FEM solution for
several decades of variation of the dimensionless parameter $R^2
p/D$.}
\label{fig:FEM2}
\end{figure}

\begin{figure}
\begin{center}
\includegraphics[width=\textwidth]{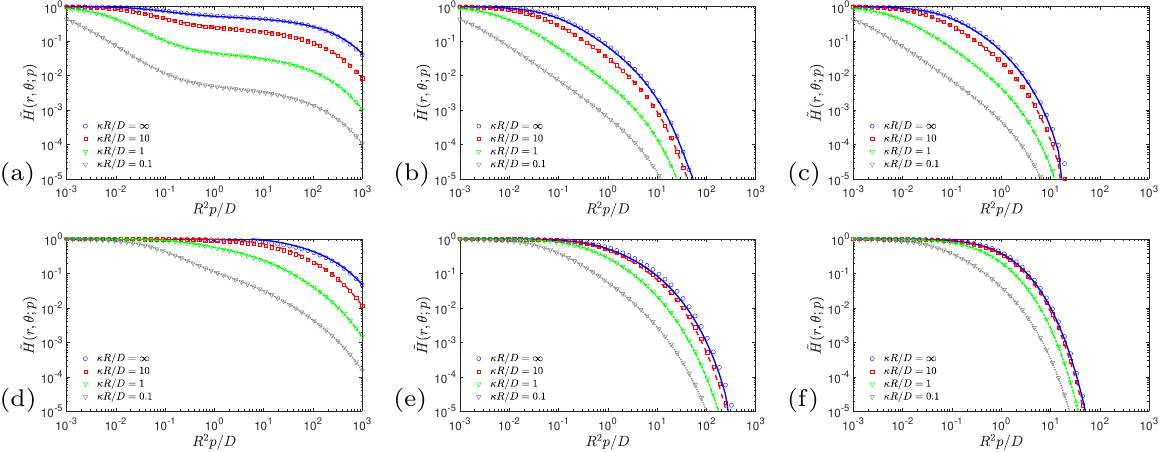}
\end{center}
\caption{\clr
Laplace-transformed PDF $\tilde{H}(r,\theta;p)$ of the FRT through a
spherical cap, with $R = 1$, $D = 1$, $r = 0.9$, $\ve = 0.1$ (top
panels) and $\ve = 1$ (bottom panels), four values of $\kappa$, and
{\bf (a,d)} $\theta = 0$, {\bf (b,e)} $\theta = \pi/2$, {\bf (c,f)}
$\theta = \pi$ (these starting points correspond to those used in
figures \ref{fig:fig22}, \ref{fig:Ht_r09_eps01_short} and
\ref{fig:Ht_r09_eps01_app}). The self-consistent approximation (lines),
truncated to $n = 50$ terms, is compared to the FEM solution of the
original problem (symbols), with $h_{\rm max} = 0.01$.  Note that our
theoretical predictions are almost indistinguishable from the FEM
solution for several decades of variation of the dimensionless
parameter $R^2 p/D$.}
\label{fig:FEMnew}
\end{figure}

In order to assess the accuracy of the self-consistent approximation,
we solve numerically the original boundary value problem for the
modified Helmholtz equation
\begin{equation}
(p - D\Delta) \tilde{H}(r,\theta;p) = 0 ,
\end{equation}
subject to the mixed boundary condition:
\begin{eqnarray}
 \frac{D}{\kappa} \partial_n \tilde H + \tilde H &=& 1   \quad (0 \leq \theta < \ve), \\
 \partial_n \tilde H &=& 0 \quad (\ve \leq \theta \leq \pi).
\end{eqnarray}
Setting $u(\rho,\theta;s) = \tilde H(\rho R,\theta; p)$ with rescaled
radial coordinate $\rho = r/R$, this equation can be written in
spherical coordinates as
\begin{equation}
\partial_\rho \rho^2 \sin\theta \partial_\rho u + \partial_\theta \sin\theta \partial_\theta u - s \rho^2 \sin \theta u = 0 ,
\end{equation}
where $s = pR^2/D$ and we omitted the azimuthal angle due to the axial
symmetry of the problem.  We solve the above equation on the planar
computational domain $C = (0,1) \times (0,\pi)$ by using a Finite
Elements Method (FEM) implemented as PDE toolbox in Matlab.  For this
purpose, we rewrite this equation in the conventional matrix form:
\begin{equation}
- \left(\begin{array}{c} \partial_\rho \\ \partial_\theta \\ \end{array}\right)^\dagger c 
\left(\begin{array}{c} \partial_\rho \\ \partial_\theta \\ \end{array}\right) u + a u = 0 ,
\end{equation}
where $c$ is the diagonal $2\times 2$ matrix with elements $\rho^2
\sin\theta$ and $\sin \theta$, and $a = s \rho^2 \sin\theta$.  This
equation is to be solved subject to the boundary conditions:
\begin{eqnarray}
&& (\partial_\rho u)_{\rho = 0} = (\partial_\theta u)_{\theta = 0} = (\partial_\theta u)_{\theta = \pi} = 0 ,\\
&& \frac{D}{\kappa R} (\partial_\rho u)_{\rho = 1} + u_{\rho = 1} = 1  \quad (0 \leq \theta < \ve) , \\
&& (\partial_\rho u)_{\rho = 1} = 0  \quad (\ve \leq \theta \leq \pi) .
\end{eqnarray}
Once the solution is found, we also evaluate the volume average
$\overline{\tilde{H}(p)}$ (by numerical integration over the
triangular mesh), and the surface average $\overline{\tilde{H}(p)}_r$
(by interpolating the solution to a line $\rho = const$ and then
integrating numerically over $\theta$).  We selected the maximal mesh
size to be $h_{\rm max} = 0.01$.  In order to verify the accuracy of
the FEM solution, we solved the problem twice, with two mesh sizes
$h_{\rm max}$ and $2h_{\rm max}$, and checked that the resulting
solutions were barely distinguishable.

Figures \ref{fig:FEM} and \ref{fig:FEM2} compare the volume-averaged
and surface-averaged Laplace-transformed PDFs
$\overline{\tilde{H}(p)}$ and $\overline{\tilde{H}(p)}_r$, obtained
via our self-consistent approximation (lines) and by FEM (symbols).
One can see that the self-consistent approximation is remarkably
accurate over a broad range of $p$, for both narrow $\ve = 0.1$ and
extended $\ve = 1$ target region.  The accuracy is higher for less
reactive targets.

{\clr 
Figure \ref{fig:FEMnew} compares the Laplace-transformed PDF
$\tilde{H}(r,\theta;p)$, obtained via our self-consistent
approximation (lines) and by FEM (symbols), for the three starting points
used in figures \ref{fig:fig22}, \ref{fig:Ht_r09_eps01_short} and
\ref{fig:Ht_r09_eps01_app}.  One can see that the self-consistent
approximation is remarkably accurate over a broad range of $p$, for
both narrow $\ve = 0.1$ and extended $\ve = 1$ target region.  The
accuracy is higher for less reactive targets.  Minor deviations seen
in panel {\bf (c)} for the case $\kappa = \infty$ are caused by the
truncation to $n = 50$ terms and can be corrected by increasing $n$
(not shown).  In turn, minor deviations in panel {\bf (e)} cannot be
corrected by increasing $n$ and can thus be attributed to a limited
accuracy of the self-consistent approximation for extended targets. }


\end{document}